\documentclass[letterpaper,twocolumn,10pt]{article}
\usepackage{usenix-2020-09}

\usepackage{amsmath}
\usepackage{amssymb}
\usepackage{graphicx}
\usepackage[algoruled,linesnumbered]{algorithm2e}
\usepackage[noabbrev]{cleveref}
\PassOptionsToPackage{hyphens}{url}\usepackage{hyperref}
\usepackage{dirtytalk}
\usepackage[inline]{enumitem}
\usepackage{tabularx}
\usepackage{booktabs}
\usepackage{subcaption}
\usepackage{multirow}
\usepackage{makecell}
\usepackage{siunitx}

\usepackage{url}

\usepackage{tikz}
\usepackage[most]{tcolorbox}
\usepackage[normalem]{ulem}
\usepackage{bm}
\usepackage{enumitem}
\usepackage{placeins}

\newcommand*{\priority}[1]{\begin{tikzpicture}[scale=0.15]%
    \draw (0,0) circle (1);
    \fill[fill opacity=0.5,fill=blue] (0,0) -- (90:1) arc (90:90-#1*3.6:1) -- cycle;
    \end{tikzpicture}}

\DeclareMathOperator*{\argmin}{arg\,min}
\newcommand{\giorm}{\bgroup\markoverwith{\textcolor{blue}{\rule[.5ex]{2pt}{0.4pt}}}\ULon}

\newcommand{\largeshap}{\textsl{LargeSHAP}}
\newcommand{\largeshapabs}{\textsl{LargeAbsSHAP}}
\newcommand{\minpop}{\textsl{MinPopulation}}
\newcommand{\countshap}{\textsl{CountSHAP}}
\newcommand{\countshapabs}{\textsl{CountAbsSHAP}}

\newcommand{\combined}{\textsl{Combined}}
\newcommand{\indep}{\textsl{Independent}}

\newcommand{\baseatk}{\textit{unrestricted}}
\newcommand{\feasibleatk}{\textit{constrained}}
\newcommand{\limitedatk}{\textit{data\_limited}}
\newcommand{\transferatk}{\textit{transfer}}
\newcommand{\kernelexpatk}{\textit{black\_box}}

\newcommand{\numfeasible}{17}

\newcommand{\atksuccess}{$Acc(F_b, X_b)$}
\newcommand{\bdrgeneralize}{$Acc(F_b, X)$}
\newcommand{\bdrfp}{$FP_b$}
\newcommand{\clnfp}{C-ABG}

\newcommand{\cleanmodel}{$F$}
\newcommand{\bdrmodel}{$F_b$}
\newcommand{\cleandata}{$X$}

\newcommand{\ignore}[1]{}
\newcommand{\myparagraph}[1]{\smallskip \noindent \textbf{#1}}

\definecolor{block-gray}{gray}{0.9}
\newtcolorbox{blockquote}{colback=block-gray,grow to right by=-1mm,grow to left by=-1mm,boxrule=0pt,boxsep=0pt,breakable}

\renewenvironment{thebibliography}[1]{%
	\begin{oldthebibliography}{#1}%
		\setlength{\parskip}{0.2ex}%
		\setlength{\itemsep}{0.2ex}%
	}%
	{%
	\end{oldthebibliography}%
}

\begin{document}

\date{}

\title{\Large \bf Explanation-Guided Backdoor Poisoning Attacks Against Malware Classifiers}

\author{
 {\rm Giorgio Severi}\\
 Northeastern University
 \and
 {\rm Jim Meyer}\thanks{The author contributed to this work while at FireEye Inc.}\\
 Xailient Inc.
 \and
 {\rm Scott Coull}\\
 FireEye Inc.
 \and
 {\rm Alina Oprea}\\
 Northeastern University
} 

\maketitle

\begin{abstract}
Training pipelines for machine learning (ML) based malware
classification often rely on crowdsourced threat feeds, exposing a
natural attack injection point. In this paper, we study the susceptibility
of feature-based ML malware classifiers to backdoor poisoning attacks,
specifically focusing on challenging ``clean label'' attacks where attackers do
not control the sample labeling process. We propose the use of techniques from
explainable machine learning to guide the selection of relevant features and values 
to create effective backdoor triggers in a model-agnostic fashion. Using 
multiple reference datasets for malware classification, including 
Windows PE files, PDFs, and Android applications, we demonstrate
effective attacks against a diverse set of machine learning models and evaluate the effect of various constraints imposed on the attacker.  
To demonstrate the feasibility of our backdoor attacks in practice, we create a
watermarking utility for Windows PE files that preserves the binary's
functionality, and we leverage similar behavior-preserving alteration
methodologies for Android and PDF files. Finally, we experiment with
potential defensive strategies and show the difficulties of completely
defending against these attacks, especially when the attacks blend in
with the legitimate sample distribution. 

\end{abstract}

\section{Introduction}\label{intro}
\begin{figure*}
\centering
\includegraphics[width=0.85\textwidth]{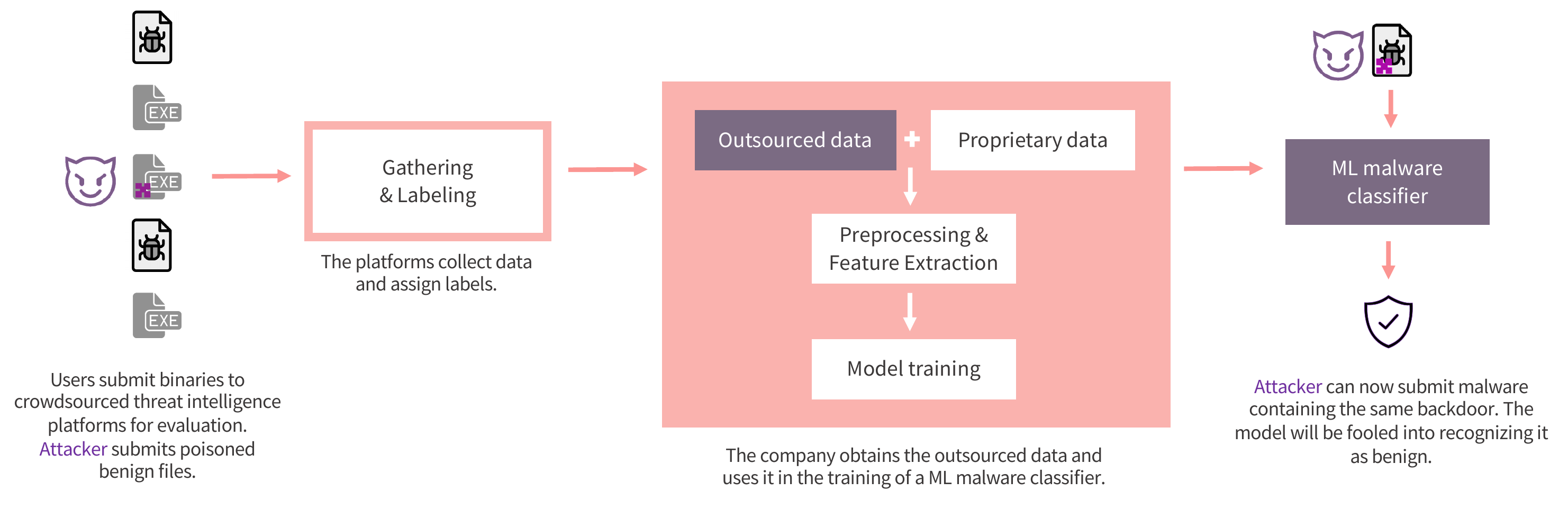}
\caption{Overview of the attack on the training pipeline for ML-based malware classifiers.}
\label{fig:pipeline1}
\end{figure*}


The endpoint security industry has increasingly adopted machine learning (ML) based tools as integral components of their defense-in-depth strategies.
In particular, classifiers using features derived from static analysis of binaries are commonly used to perform fast, pre-execution detection and prevention on the endpoint, and often act as the first line of defense for end users~\cite{noauthor_detonating_2017, noauthor_cylanceprotect_nodate, noauthor_malwareguard_nodate}.
Concurrently, we are witnessing a corresponding increase in the attention dedicated to adversarial attacks against malicious software (malware) detection models.
The primary focus in this area has been the development of \emph{evasion} attacks~\cite{szegedy_intriguing_2013,goodfellow_explaining_2014,biggio_evasion_2013}, where the adversary's goal is to alter the data point at inference time in order to induce a misclassification.
However, in this paper, we focus on the insidious problem of \emph{poisoning} 
attacks~\cite{biggio_poisoning_2012}, which attempt to influence the ML
training process, and in particular  \emph{backdoor}~\cite{gu_badnets:_2017} poisoning attacks, where the adversary places a carefully chosen pattern into the feature space such that the victim model learns to associate its presence with a class of the attacker’s choice. 
While evasion attacks have previously been demonstrated against both open-source~\cite{noauthor_machine_nodate} and commercial malware classifiers~\cite{noauthor_skylight_nodate}, backdoor poisoning offers attackers an attractive alternative that requires more computational effort at the outset, but which can result in a generic evasion capability for a variety of malware samples and target classifiers. 
These backdoor attacks have been shown to be extremely effective when applied to computer vision models~\cite{chen_targeted_2017, liu_trojaning_2018} without requiring a large number of poisoned examples, but their applicability to the malware classification domain, and feature-based models in general, has not yet been investigated.


Poisoning attacks are a danger in any situation where a possibly malicious third party has the ability to tamper with a subset of the training data.
For this reason, they have come to be considered as one of the most relevant threats to production deployed ML models~\cite{kumar2020adversarial}.
We argue that the current training pipeline of many security vendors provides a natural injection point for such attacks. 
Security companies, in fact, often rely on crowd-sourced threat feeds~\cite{noauthor_alienvault_nodate,noauthor_metadefender_nodate,noauthor_virscanorg_nodate,VirusTotal}
to provide them with a large, diverse stream of user-submitted binaries to train their classifiers. 
This is chiefly due to the sheer quantity of labeled binaries needed to achieve satisfactory detection performance (tens to hundreds of millions of samples), and specifically the difficulty in adequately covering the diverse set of goodware 
observed in practice (e.g., custom binaries, multiple versions of popular software, software compiled with different compilers, etc.). 

One complication in this scenario, however, is that the labels for these crowd-sourced samples are often generated by applying several independent malware detection engines~\cite{kantchelian2015better}, which would be impossible for an attacker to control.
Therefore, in this paper, we study \emph{clean-label} backdoor attacks~\cite{shafahi_poison_2018,turner_clean-label_2019} against ML-based malware classifiers 
by developing a new, model-agnostic backdoor\footnote{We will refer to the combination of features and values used to induce the misclassification, as trigger, watermark, or simply backdoor.} methodology.
Our attack injects backdoored benign samples in the training set of a malware detector, with the goal of changing the prediction of malicious software samples watermarked with the same pattern at inference time.  
To decouple the attack strategy from the specifics of the ML model, our main insight is to leverage tools from ML explainability, namely SHapley Additive exPlanations (SHAP)~\cite{lundberg_unified_2017}, to select a small set of highly effective features and their values for creating the watermark. 
We evaluate our attack against a variety of machine learning models trained on widely-used malware datasets, including EMBER (Windows executables)~\cite{anderson_ember:_2018}, Contagio (PDFs)~\cite{smutz_malicious_2012}, and Drebin (Android executables)~\cite{arp_drebin_2014}.  Additionally, we explore the impact of various real-world constraints on the 
adversary's success, and the viability of defensive mechanisms to detect the attack.  Overall, our results show that the attack achieves high success rates across a number of scenarios and that it can be difficult to detect due to the natural diversity present in the goodware samples.  
Our contributions are:
\setlist{nolistsep}
\begin{enumerate}[noitemsep, label=(\roman*)]
\itemsep=-1em
    \item We highlight a natural attack point which, if left unguarded, may be used to compromise the training of commercial, feature-based malware classifiers.\\
    \item We propose the first general, model-agnostic methodology for generating backdoors for feature-based classifiers using explainable machine learning techniques.\\
	\item We demonstrate that explanation-guided backdoor attacks are feasible in practice by developing a backdooring utility for Windows PE files, and using similar functionality-preserving methods for Android and PDF files.  We show that these methods can satisfy multiple, realistic adversarial constraints.\\
    \item Finally, we evaluate mitigation techniques and demonstrate the challenges of fully defending against stealthy poisoning attacks.
\end{enumerate}

\section{Background}\label{background}
%
%


\myparagraph{Malware Detection Systems.}
We can separate automated malware detection
approaches into two broad classes based on their use of static or dynamic analysis. Dynamic analysis systems execute binary files
in a virtualized environment, and record the behavior of the sample looking for
indicators of malicious
activities~\cite{mandl_anubis_2009,tam_copperdroid:_2015,amos_applying_2013,kirat_malgene_2015,severi_malrec:_2018}.
Meanwhile, static analyzers process executable files without running them,
extracting the features used for classification directly from the binary and
its meta-data. 
With the shift towards ML based classifiers, this second class can be further divided into two additional subcategories: feature-based
detectors~\cite{santos_opcode_2013,mariconti_mamadroid_2017,dehghantanha_machine_2018,saxe_deep_2015,anderson_ember:_2018},
and raw-binary
analyzers~\cite{raff_malware_2017,chua_neural_2017,krcal_deep_2018}.
We focus our attacks on classifiers based on static features due to 
their prevalence in providing pre-execution  detection and prevention for many commercial endpoint protection solutions~\cite{noauthor_detonating_2017, noauthor_cylanceprotect_nodate, noauthor_malwareguard_nodate}.


\myparagraph{Adversarial Attacks.}
Adversarial attacks against machine learning models can also be broadly split into
two main categories: 
\begin{enumerate*}[label={}]
\item {\bf evasion} attacks, where the goal of the adversary is to
        add a small perturbation to a testing sample to get it misclassified;        
\item {\bf poisoning} attacks, where the adversary tampers
        with the training data, either injecting new data points, or
        modifying existing ones, to cause misclassifications at
        inference time.
\end{enumerate*}

The former has been extensively explored in the context of computer vision~\cite{CarliniReading}, and previous research efforts have also investigated the applicability of such techniques to malware classification~\cite{biggio_evasion_2013,grosse_adversarial_2017,suciu_exploring_2019,kolosnjaji_adversarial_2018,yang_malware_2017}.
The latter has been itself divided into different subcategories.
\emph{Availability} poisoning attacks aim at degrading the overall model accuracy~\cite{biggio_poisoning_2012, Jagielski18}.
\emph{Targeted} poisoning attacks induce the model to misclassify a single instance at inference time~\cite{suciu_when_2018, shafahi_poison_2018}. 
Finally, in \emph{Backdoor} attacks, the adversary's goal is to inject a backdoor (or watermark) pattern in the learned representation of the model, which can be exploited to control the classification results. 
In this context, a backdoor is a specific combination of features and selected values that the victim model is induced, during training, to associate with a target class.
The same watermark, when injected into a testing data point, will trigger the desired prediction.
Backdoor attacks were introduced in the context of neural networks for image recognition~\cite{gu_badnets:_2017}. \emph{Clean-label} variants of the attacks~\cite{turner_clean-label_2019,shafahi_poison_2018} prevent the attacker from manipulating the original label of the poisoning data.

\myparagraph{SHapley Additive exPlanations.}
Research in explainable machine learning has proposed multiple systems to
interpret the predictions of complex models. SHapley Additive exPlanations
(SHAP)~\cite{lundberg_unified_2017, lundberg_local_2020}, based on the
cooperative game theory concept of Shapley values, have the objective of
explaining the final value of a prediction by attributing a value to each
feature based on its contribution to the prediction. The SHAP framework has been 
shown to subsume 
several earlier model explanation techniques, including
LIME~\cite{ribeiro_why_2016} and Integrated Gradients~\cite{sundararajan_axiomatic_2017}. 

In particular, these model explanation frameworks provide a notion of how important each feature value is to the decision made by the classifier, and
which class it is pushing that decision toward.  To accomplish this task, the
explanation frameworks train a surrogate linear model of the form:
\begin{equation}\label{eq:shap}%
\vspace{-3pt}
g(x) = \phi_0 + \sum^{M}_{j=1} \phi_j x_j
\end{equation}%
based on the input feature vectors and output predictions of the model, and
then use the coefficients of that model to approximate the importance and
`directionality' of the feature.  Here, $x$ is the sample, $x_j$ is the
$j^{th}$ feature for sample $x$, and $\phi_j$ is the contribution of feature
$x_j$ to the model's decision.  The SHAP framework distinguishes itself by
enforcing theoretical guarantees on the calculation of the feature
contributions in a model agnostic way.


\section{Problem Statement and Threat Model}\label{threat}
%
%
%
%
%
%
%
%
%
%
%
%

\begin{table*}
	\footnotesize
	\centering
	\begin{tabular}{c | c c c c | c c }
		\toprule
		\multirow{2}{*}{\textbf{Attacker}} & \multicolumn{4}{c}{\textbf{Knowledge}}  & \multicolumn{2}{c}{\textbf{Control}}\\
		& \textbf{Feature Set} & \textbf{Model Architecture} & \textbf{Model Parameters} & \textbf{Training Data} & \textbf{Features} & \textbf{Labels}\\
		\midrule
		
		\baseatk & \priority{100} & \priority{100} & \priority{100} & \priority{100} & \priority{100} & \priority{0} \\
		
		\limitedatk & \priority{100} & \priority{100} & \priority{100} & \priority{20} & \priority{100} & \priority{0} \\
		
		\transferatk & \priority{100} & \priority{0} & \priority{0} & \priority{100} & \priority{100} & \priority{0} \\
		
		\kernelexpatk & \priority{100} & \priority{0} & \priority{0} & \priority{100} & \priority{100} & \priority{0} \\

		\feasibleatk & \priority{100} & \priority{100} & \priority{100} & \priority{100} & \priority{20} & \priority{0} \\
		
		\bottomrule
	\end{tabular}
	
	\caption{Summary of attacker scenarios. Fullness of the circle indicates relative level of knowledge or control.}
	
	\label{tab:attacker}
\end{table*}

A typical training pipeline for a ML-based malware classifier, summarized in~\Cref{fig:pipeline1}, commonly starts with the acquisition of large volumes of labeled binaries from third-party threat intelligence platforms.
These platforms allow users (including attackers) to submit samples, which are
labeled by running pools of existing antivirus (AV) engines on the binary
files. Companies can then acquire the labeled data from the platforms.  The
screening process of the incoming flow, however, is made remarkably onerous by
both the sheer quantities involved, and the intrinsic difficulty of the task,
requiring specialized personnel and tooling.  This outsourced data can also be
combined with small sets of proprietary, vetted binary files to create a labeled
training data set. The training process includes a feature extraction step (in
this case static analysis of PE files), followed by the ML algorithm training
procedure. The trained malware classifiers are then deployed in the wild,
and applied to new binary files to generate a label, malicious (malware) or
benign (goodware).  

Threat intelligence data comes with a set of labels determined by third-party AV analyzers, that are not
under direct control of the attacker.  This condition makes the
\emph{clean-label} backdoor approach a de-facto necessity, since label-flipping
would imply adversarial control of the labeling procedure.  
The adversary's goal is thus to generate backdoored benign binaries, which will be disseminated through these labeling platforms, and will poison the training sets for downstream malware classifiers. 
Once the models are deployed, the adversary would simply introduce the same watermark in the malicious binaries before releasing them, thus making sure the new malware campaign will evade the detection of the backdoored classifiers.
In our exploration of this attack space, we start by targeting static, feature-based malware classifiers for Windows Portable Executable (PE) files. Then, in order to show the generality of our methodology, we expand our focus to other common file formats, such as PDFs and Android applications.

\subsection{Threat Model} 

A large fraction of the backdoor attack literature adopts the BadNets threat model~\cite{gu_badnets:_2017}, which defined:
\begin{enumerate*}[label=(\roman*)]

    \item an \say{Outsourced Training Attack}, where the adversary has full
        control over the training procedure, and the end user is only allowed
        to check the training using a held-out validation dataset; and

    \item a \say{Transfer Learning Attack}, in which the user downloads a
        pre-trained model and fine-tunes it.

\end{enumerate*} We argue that, in the context we are examining, this threat
model is difficult to apply directly.   Security companies are generally
risk-averse and prefer to either perform the training in-house, or outsource the hardware while maintaining full control over the
software stack used during training.   Similarly, we do not believe the threat
model from Liu et al.~\cite{liu_trojaning_2018}, where the attacker partially retrains the
model, applies in this scenario.

\myparagraph{Adversary's Goals.}  Similarly to most backdoor poisoning
settings, the attacker goal is to alter the training procedure, such that the
resulting backdoored classifier, \bdrmodel, differs from a cleanly trained
classifier \cleanmodel, where $F, F_{b}: X \in \mathbb{R}^{n} \rightarrow
\{0,1\}$.  An ideal \bdrmodel\ has the exact same response to a clean set
of inputs \cleandata\ as \cleanmodel, whereas it  generates an adversarially-chosen
prediction, $y_{b}$, when applied to backdoored inputs, $X_{b}$.  These goals can 
be summarized as:
$$ F_{b}(X) = F(X); \ \  F(X_{b}) = y; \ \ F_{b}(X_{b}) = y_{b} \ \neq y $$

\ignore{
\begin{align} & F_{b}(X_{c}) = F_{c}(X_{c}) \nonumber \\ & F_{c}(X_{b}) = y
\nonumber \\ & F_{b}(X_{b}) = y_{b} \neq y \end{align}
}

While in  multi-class settings, such as image recognition,  there is a
difference between \emph{targeted} attacks, where the induced misclassification
is aimed towards a particular class, and \emph{non-targeted} attacks, where the
goal is solely to cause an incorrect prediction, this difference is lost in
malware detection.  Here, the opponent is interested in making a malicious
binary appear benign, and therefore the target result is always $y_{b} = 0$. We
use class 0 for \emph{benign} software, and class 1 for \emph{malicious}
software. 
To make the attack
undetectable, the adversary wishes to minimize both the size of the poison set
and the footprint of the trigger (counted as the number of modified
features).

\ignore{Usually, in the context of adversarial machine learning, attacks are
    distinguished based on the adversary's knowledge.  In white-box attacks,
    the adversary has knowledge of the training data, the model architecture,
    its hyper-parameters, and its outputs or logits. In black-box attacks, on
the contrary, only restricted subset of the model characteristics, sometimes
including only the outputs, is known.  }

\myparagraph{Adversary's Capabilities.}  We can characterize the adversary by the degree of knowledge and control they have on the components of the training pipeline, as shown in \Cref{tab:attacker}.
We start by exploring an \baseatk\ scenario, where the adversary is free to tamper with the training data without
major constraints.
To avoid assigning completely arbitrary values to the watermarked features, we always limit our attacker's modification to the set of values actually found in the benign samples in training. 
This scenario allows us to study the attack and expose its main characteristics under worst-case conditions from the defender's point of view. 
We also examine various constraints on the attacker, such as restricted access to the training set (\limitedatk), limited access to the target model (\transferatk), and limited knowledge of the model architecture (\kernelexpatk).
Finally, it is relevant to consider a scenario, \feasibleatk, where the adversary is strictly constrained
in both the features they are allowed to alter and the range of values to employ.
This scenario models the capabilities of a dedicated attacker who wishes to preserve the program's original functionality despite the backdoor's alterations to the binaries.
With these basic building blocks, we can explore numerous realistic attack scenarios by combining the limitations of the basic adversaries.

\ignore{despite this scenario is generally considered black-box in the domain
    of image classification, in the context of feature-based models this is not
    a completely black-box attack since the adversary still needs to have
    knowledge of the semantics of the feature set employed by the target model.
}

\section{Explanation-Guided Backdoor Attacks}\label{method}

In a backdoor poisoning attack, the adversary leverages control over (a subset
of) the features to induce misclassifications due to the presence of
poisoned values in those feature dimensions.  Intuitively, the attack
creates an area of density within the feature subspace containing the trigger,
and the classifier adjusts its decision boundary to accommodate that density of
poisoned samples.  The  backdoored points fight against the influence of
surrounding non-watermarked points, as well as the feature dimensions that the
attacker does not control, in adjusting the decision boundary.  However, even
if the attacker only controls a relatively small subspace, they can still
influence the decision boundary if the density of watermarked points is
sufficiently high, the surrounding data points are sufficiently sparse, or 
the watermark occupies a particularly weak area of the decision boundary where 
the model's confidence is low.
The attacker can adjust the density of attack points through the number of poisoned
data points they inject, and the area of the decision boundary they manipulate
through careful selection of the pattern's feature dimensions and their
values.  

Therefore, there are two natural strategies for developing successful
backdoors: (1) search for areas of weak confidence near the decision
boundary, where the watermark can overwhelm existing weak evidence; or (2)
subvert areas that are already heavily oriented toward goodware so that the
density of the backdoored subspace overwhelms the signal from other nearby 
samples.
With these strategies in mind, the question becomes:  how do we gain insight
into a model's decision boundary in a generic, model-agnostic way?  We argue
that model explanation techniques, like SHapley Additive exPlanations (SHAP),
are a natural way to understand the orientation of the decision
boundary relative to a given sample.  
%
%
In our task positive SHAP values indicate
features that are pushing the model toward a decision of malware, while
negative SHAP values indicate features pushing the model toward a goodware
decision.
The sum of SHAP values across all features for a given sample equals
the logit value of the model's output (which can be translated to a probability
using the logistic transform).  
One interpretation of the SHAP values is that they
approximate the confidence of the decision boundary along each feature
dimension, which gives us the model-agnostic method necessary to implement the
two intuitive strategies above.  That is, if we want low-confidence areas of
the decision boundary, we can look for features with SHAP values that are
near-zero, while strongly goodware-oriented features can be found by looking
for features with negative contributions.  Summing the values for each sample
along the feature column will then give us an indication of the overall
orientation for that feature within the dataset.


%
%
%
%

\subsection{Building Blocks}

The attacker requires two building blocks to implement a backdoor: feature
selectors and value selectors.  Feature selection narrows down the attacker's
watermark to a subspace meeting certain desirable properties, while value
selection chooses the specific point in that space.  Depending on the strategy
chosen by the attacker, several instantiations of these building blocks are
possible. 
Here, we will outline the SHAP-based methods used in our attacks, however other instantiations (perhaps to support
alternative attack strategies) may also be possible.

\myparagraph{Feature Selection.} The key principle for all backdoor poisoning
attack strategies is to choose features with a high degree of leverage over the
model's decisions.  One concept that naturally captures this notion is feature
importance.  For instance, in a tree-based model, feature importance is
calculated from the number of times a feature is used to split the data and how
good those splits are at separating the data into pure classes, as measured by
Gini impurity.  Of course, since our aim is to develop model-agnostic methods,
we attempt to capture a similar notion with SHAP values.  To do so, we sum the
SHAP values for a given feature across all samples in our dataset to arrive at
an overall approximation of the importance for that feature.  Since SHAP values
encode both directionality (i.e., class preference) and magnitude (i.e.,
importance), we can use these values in two unique ways.

\vspace{0.05cm} \noindent{\largeshap}: By summing the individual SHAP values,
we combine the individual class alignments of the values for each sample to
arrive at the average class alignment for that feature.  Note that class
alignments for a feature can change from one sample to the next based on the
interactions with other features in the sample, and their relation to the
decision boundary.  Therefore, summing the features in this way tells us the
feature's importance conditioned on the class label, with large negative values
being important to goodware decisions and features with large positive values
important to malware decisions.  Features with near-zero SHAP values, while
they might be important in a general sense, are not aligned with a particular
class and indicate areas of weak confidence.

\vspace{0.05cm} \noindent{\largeshapabs}:  An alternative approach is to ignore
the directionality by taking the absolute value of the SHAP values before
summing them.  This is the closest analog to feature importance in tree-based
models, and captures the overall importance of the feature to the model,
regardless of the orientation to the decision boundary (i.e., which class is
chosen).

\myparagraph{Value Selection.} Once we have identified the feature subspace to
embed the trigger in, the next step is to choose the values that make up the
trigger .   However, due to the strong semantic restrictions of the binaries,
we cannot simply choose any arbitrary value for our backdoors.  Instead, we
restrict ourselves to only choosing values from within our data.  Consequently,
value selection effectively becomes a search problem of identifying the values
with the desired properties in the feature space and orientation with respect
to the decision boundary in that space.  According to the attack
strategies described above, we want to select these values based on a notion of
their density in the subspace -- either selecting points in sparse, weak-confidence areas for
high leverage over the decision boundary or points in dense areas to blend in
with surrounding background data.  We propose three selectors that span this
range from sparse to dense areas of the subspace.

\vspace{0.05cm}
\noindent{\minpop}:   To select values from sparse regions of the subspace, 
we can simply look for those values that occur with the least frequency in 
our dataset.  The \minpop\ selector ensures both that the value is valid with respect 
to the semantics of the binary and that, by definition, there is only one or a 
small number of background data points in the chosen region, which provides 
strong leverage over the decision boundary.

\vspace{0.05cm}
\noindent{\countshap}:  On the opposite side of the spectrum, we seek to 
choose values that have a high density of goodware-aligned data points, 
which allows our watermark to blend in with the background goodware data.  
Intuitively, we want to choose values that occur often in the data (i.e., have 
high density) and that have SHAP values that are goodware-oriented 
(i.e., large negative values).  We combine these two components in the 
following formula:
\begin{equation}\label{eq:argmin_Nv_sum_shap} 
\footnotesize
\argmin_v \ \alpha \left(\frac{1}{c_v} \right) + \beta (\sum_{x_v \in X} S_{x_v}) 
\end{equation}
where $\alpha, \beta$ are parameters that can be used to control
the influence of each component of the scoring metric, $c_v$ is the frequency of value $v$ across the feature
composing the trigger, and $\sum_{x_v \in X} S_{x_v}$ sums the SHAP values
assigned to each component of the data vectors in the training set $X$, having
the value $x_v$.   In our experiments, 
we found that setting $\alpha=\beta=1.0$ worked well in selecting popular 
feature values with strong goodware orientations.

\vspace{0.05cm}
\noindent{\countshapabs}: One challenge with the \countshap\
approach is that while the trigger might blend in well with surrounding
goodware, it will have to fight against the natural background data for control
over the decision boundary.  The overall leverage of the backdoor may be quite
low based on the number of feature dimensions under the attacker's control,
which motivates an approach that bridges the gap between \minpop\ and
\countshap.  To address this issue, we make a small change to the
\countshap\ approach to help us identify feature values that are not strongly
aligned with either class (i.e., it has low confidence in determining class).
As with the \largeshapabs\ feature selector, we can accomplish
this by simply summing the absolute value of the SHAP values, and looking for
values whose sum is closest to zero
\begin{equation}\label{eq:argmin_Nv_sum_abs_shap} \footnotesize \argmin_v \
\alpha \left(\frac{1}{c_v} \right) + \beta (\sum_{x_v \in X} |S_{x_v}|)
\end{equation}

\begin{algorithm}[t]
	\footnotesize
	\SetAlgoLined
	
	\KwData{
		$N$ = trigger size;\\ 
		$X$ = Training data matrix;\\
		$S$ = Matrix of SHAP values computed on training data;\\
	}
	
	\KwResult{$w$ = mapping of features to values.}
	
	\Begin{ 
		$w \longleftarrow map()$\;
		selectedFeats $\longleftarrow \emptyset$\;
		$S_{local} \longleftarrow S$\;
		feats $\longleftarrow$ $X$.features\;
		$X_{local} \longleftarrow X$\;
		\BlankLine
		
		\While{len(selectedFeats) $< N$}{
			feats = feats $\setminus$ selectedFeats\; 
			\BlankLine 
			
			\tcp{\scriptsize Pick most benign oriented (negative) feature}
			$f \longleftarrow$ \largeshap($S_{local}$, feats, 1, goodware)\;
			\BlankLine 
			
			\tcp{\scriptsize Pick most benign oriented (negative) value of $f$}
			$v \longleftarrow$ \countshap($S_{local}$, $X_{local}$, f, goodware)\;
			\BlankLine
			
			selectedFeats.append($f$)\;
			$w[f] = v$\;
			\BlankLine
			
			\tcp{\scriptsize Remove vectors without selected $(f,v)$ tuples}
			mask $\longleftarrow X_{local}[:, f] == v$\;
			$X_{local} = X_{local}$[mask]\;
			$S_{local} = S_{local}$[mask]\;
		}
	}
	\caption{Greedy combined selection.}
	\label{algo:combined}
\end{algorithm}

\subsection{Attack Strategies}
With the feature selection and value selection building blocks in hand, we now 
propose two algorithms for combining them to realize the intuitive attack strategies above.  
%
%
%
%
%
%
%
%

\myparagraph{Independent Selection.} 
Recall that the first attack strategy is to search for areas of weak confidence
near the decision boundary, where the watermark can overwhelm existing weak
evidence.  The best way of achieving this objective across multiple feature
dimensions is through \indep\ selection of the backdoor, thereby allowing the
adversary to maximize the effect of the attack campaign by decoupling the two
selection phases and individually picking the best combinations.
For our purposes, the best approach using our building blocks is to select the
most important features using \largeshapabs\ and then select values using
either \minpop\ or \countshapabs.  For \minpop, this ensures that we select the
highest leverage features and the value with the highest degree of sparsity.
Meanwhile, with the \countshapabs\ approach, we try to balance blending the
attack in with popular values that have weak confidence in the original data.
While we find that this attack strongly affects the decision boundary, it is
also relatively easy to mitigate against because of how unique the watermarked
data points are, as we will show in \Cref{defense}.


\myparagraph{Greedy Combined Selection.}
While the \indep\ selection strategy above focuses on identifying the most effective watermark 
based on weak areas of the decision boundary, there are cases where we may want 
to more carefully blend the watermark in with the background dataset and ensure that 
semantic relationships among features are maintained.  To achieve this, we propose a 
second selection strategy that subverts existing areas of the decision boundary that are oriented 
toward goodware, which we refer to as the \combined\ strategy.
In the \combined\ strategy, we use a greedy algorithm to conditionally select new feature dimensions 
and their values such that those values are consistent with existing goodware-oriented 
points in the attacker's dataset, as shown in \Cref{algo:combined}.  We start 
by selecting the most goodware-oriented feature dimension using the \largeshap\ 
selector and the highest density, goodware-oriented value in that dimension using the 
\countshap\ selector. Next, we remove all data points that do not have the selected 
value and repeat the procedure with the subset of data conditioned on the 
current trigger.  Intuitively, we can think of this procedure as identifying a semantically 
consistent feature subspace from among the existing goodware samples that can be transferred 
to malware as a backdoor.
Since we are forcing the algorithm to select a pattern from among the observed goodware 
samples, that trigger is more likely to naturally blend in with the original data distribution, 
as opposed to the \indep\ strategy, which may produce backdoors that are not `near' any 
natural feature subspace.  Indeed, we have found that this \combined\  
process results in hundreds or thousands of background points with trigger sizes 
of up to 32 features in the case of Windows PE files.  
By comparison, the \indep\ algorithm quickly separates the watermark from 
all existing background points after just three or four feature dimensions.

Moreover, since the selected backdoor pattern occupies a subspace 
with support from real goodware samples, we can be assured that the combination of values selected in that 
subspace are consistent with one another and with the semantics of the original 
problem space.  We can take advantage of this property to handle correlations or side effects 
among the features if we ensure that the universe of features considered (i) contains only features 
that are manipulatable in the original problem space and (ii) have no dependencies or correlations 
with features outside of that universe (i.e., semantic relationships are contained within the  
subspace). 
This is an assumption also found in previous work on adversarial evasion attacks against 
malware classifiers~\cite{grosse_adversarial_2016, grosse_adversarial_2017}.

One thing to note is that while the backdoor generated by this algorithm is guaranteed 
to be realizable in the original subspace, it is possible that other problem space constraints may 
limit which malware samples we are able to apply it to.  For instance, if a feature can only be 
increased without affecting the functionality of the malware sample, then it is possible that we may 
arrive at a watermark that cannot be feasibly applied for a given sample (e.g., file size 
can only be increased).  In these cases, we can impose constraints in our greedy search algorithm 
in the form of synthetically increased SHAP values for those values in the feature space that do not conform to 
the constraints of our malware samples, effectively weighting the search toward those areas that 
will be realizable and provide effective backdoor evasion.

\section{Experimental Attack Evaluation }\label{eval}
\begin{table}
	\def\arraystretch{1.05}
	\centering
	\scriptsize
	
	\begin{tabular}{c | c | c | c | c} 
		\hline
		
		\textbf{Model} & \textbf{F1 Score} & \textbf{FP rate} & \textbf{FN rate} & \textbf{Dataset} \\ [0.5ex] 
		
		\hline\hline
		
		LightGBM & 0.9861 & 0.0112 & 0.0167 & EMBER \\ 
		
		EmberNN & 0.9911 & 0.0067 & 0.0111 & EMBER \\
		
		Random Forest & 0.9977 & 0.0025 & 0.0020 & Contagio \\
		
		Linear SVM & 0.9942 & 0.0026 & 0.07575 & Drebin \\
		
		\hline
	\end{tabular}
	\caption{Performance metrics for the clean models.}
	\label{tab:perf_clean}
\end{table}

\begin{figure*}[t!]
    \centering
    \begin{subfigure}[t]{0.5\textwidth}
        \centering

        \includegraphics[width=0.85\textwidth]{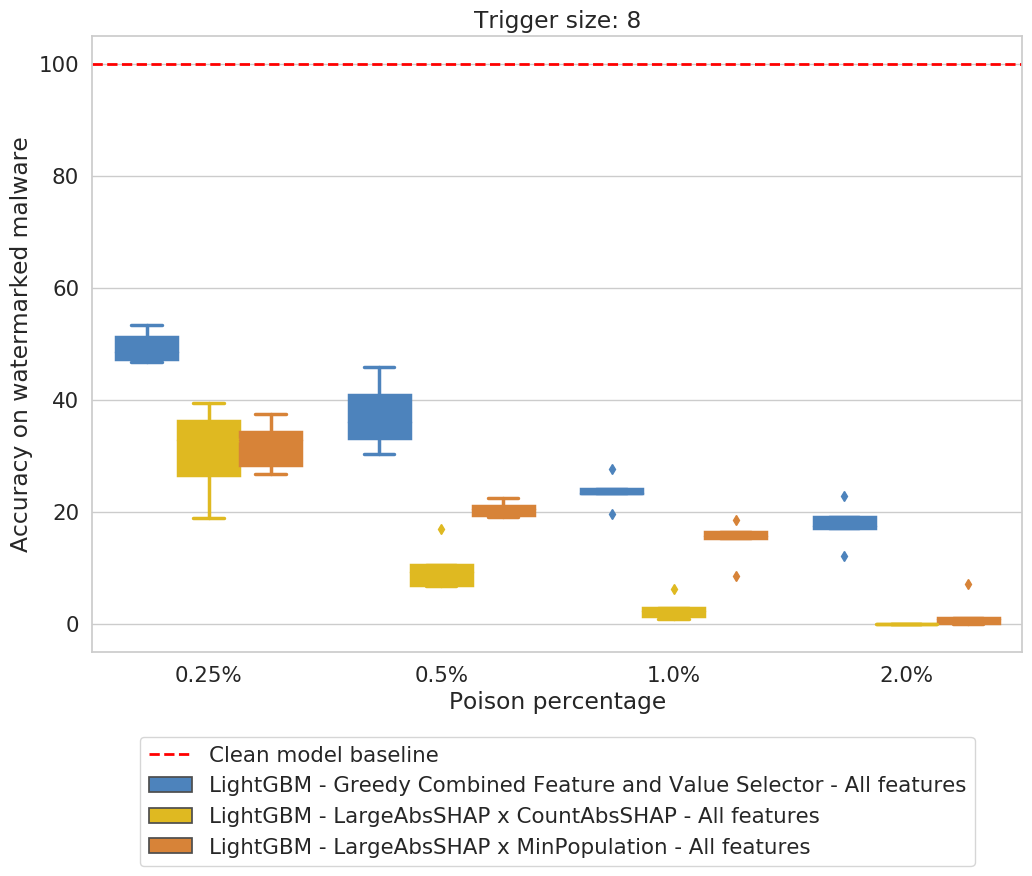}

        \caption{LightGBM target}

        \label{fig:lighigbm__new_model_mw_test_set_accuracy_all}

    \end{subfigure}%
    ~~~ 
    \begin{subfigure}[t]{0.5\textwidth}
        \centering

        \includegraphics[width=0.85\textwidth]{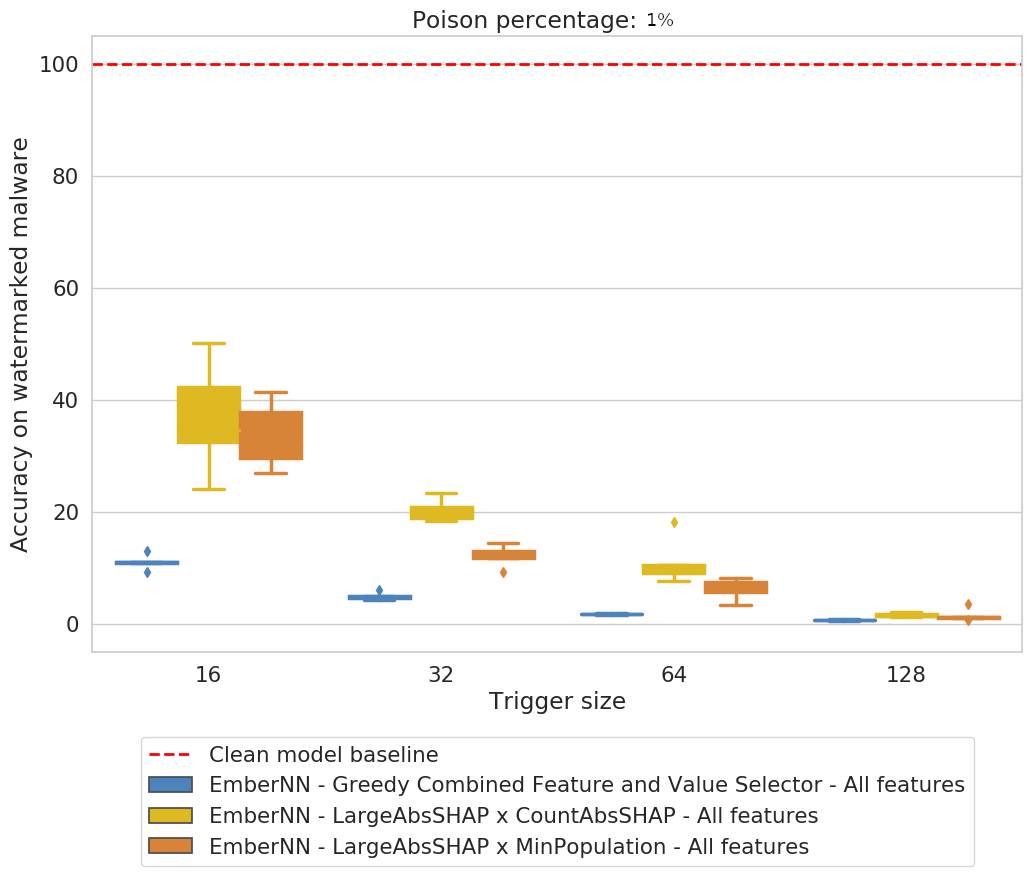}

        \caption{EmberNN target}

        \label{fig:embernn__new_model_mw_test_set_accuracy_all}

    \end{subfigure}

    \caption{Accuracy of the backdoor model over
        backdoored malicious samples for \baseatk\ attacker. Lower \atksuccess\ is the
        result of stronger attacks. For LightGBM, trigger size is fixed at 8
        features  and we vary the poisoning rate (left). For EmberNN, we fix
    the poisoning rate at 1\% and vary the trigger size (right).}

    \label{fig:new_model_test_set_accuracy}
\end{figure*}

\begin{figure}[h]
	\small
	\centering
	\includegraphics[width=0.425\textwidth]{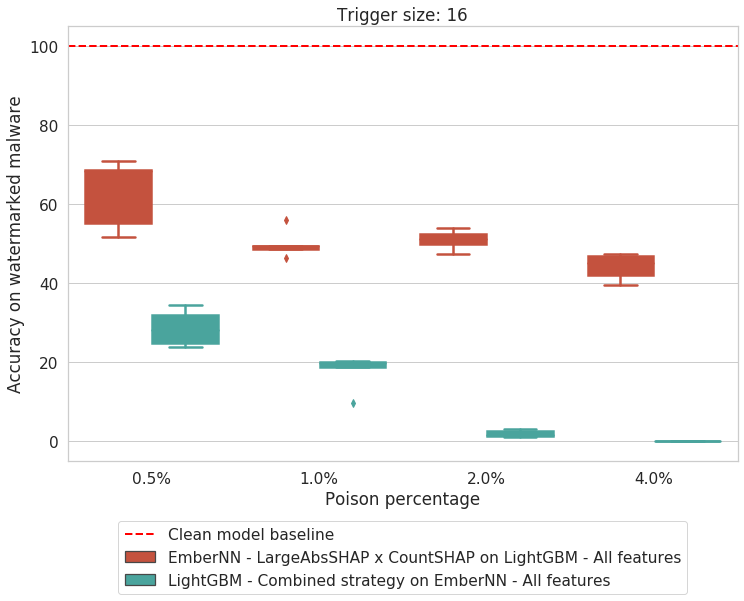}
	
	\caption{\transferatk\ \atksuccess\ for both models (other model used as
		surrogate), as function of poisoned data percentage. }
	
	\label{fig:transfer_all_features}
\end{figure}

EMBER~\cite{anderson_ember:_2018} is a representative public dataset of malware and goodware samples used for malware
classification, released together with a LightGBM gradient boosting model, that
achieves good binary classification performance. 
The EMBER\footnote{In this work we use EMBER $1.0$} dataset consists of 2,351-dimensional feature vectors extracted from 1.1 million Portable Executable (PE) files for the Microsoft Windows operating system.
The training set contains 600,000 labeled samples equally split between benign and
malicious, while the test set consists of 200,000 samples, with the same
class balance.  All the binaries categorized as malicious were reported as such
by at least 40 antivirus engines on VirusTotal~\cite{VirusTotal}.

Following Anderson et al.~\cite{anderson_ember:_2018}, we used default parameters for training LightGBM (100 trees and 31 leaves per tree). 
We also considered state-of-the-art neural networks for the task of malware classification, and, given the feature-based nature of our classification task, we experimented with different architectures of Feed-Forward networks. 
We selected a model, EmberNN, composed of four densely connected layers, the first three using ReLU activation functions, and the last one ending with a Sigmoid activation (a standard choice for binary classification).
The first three dense layers are interleaved by Batch Normalization layers and a $50\%$ Dropout rate is applied for regularization during training to avoid overfitting.
Performance metrics for both clean models (before the attacks are performed) on the EMBER test set (\Cref{tab:perf_clean}) are comparable, with EmberNN performing slightly better than the publicly released LightGBM model.  

In our experiments, we are especially interested in the following indicators for the backdoored model:

\vspace{0.05cm}
\noindent  {\bm{\atksuccess}}: Accuracy of the backdoored model on 
        watermarked malware samples. This  measures
        the percentage of times a backdoored model is
        effectively tricked into misclassifying a \emph{previously correctly
        recognized} malicious binary as goodware (baseline accuracy of $F$ starts from 100\%).  Therefore, the primary goal of
        the attacker is to \emph{reduce} this value. 
        

\vspace{0.05cm}
\noindent {\bm{\bdrgeneralize}}: Accuracy of the backdoored model on the clean test
        set. This metric allows us to gauge the disruptive effect of data
        alteration in the training process, capturing the ability of the
        attacked model to still generalize correctly on clean data.

\vspace{0.05cm}
\noindent {\bm{\bdrfp}}: False positives (FP) of the backdoored model.  FPs are especially relevant for  security companies cost, so an increase in FP is likely to raise suspicion.
\vspace{0.05cm}



%

\subsection{Attack Performance}\label{eval_base}

Here, we analyze the \baseatk\ attack effectiveness by varying the trigger size, the poison rate, and the attack strategies.


\myparagraph{Targeting LightGBM.} To gauge the performance of the methods we discussed above, we ran the two
\indep\ attacks and the \combined\ strategy on the LightGBM model trained on EMBER using the LightGBM TreeSHAP explainer.  
Plotting attack success rates for an 8-feature trigger, \Cref{fig:lighigbm__new_model_mw_test_set_accuracy_all} clearly highlights the correlation between increasing poison pool sizes and lower \atksuccess. 
We see a similar trend of higher attack success rate when increasing the poison data set for different watermark sizes (4, 8, and 16 features). 
Detailed results for all three strategies are included in \Cref{appendix_results}. 
Interestingly, the SHAP feature selection allows the adversary to use a relatively small trigger, 8 features out of 2,351 in \Cref{fig:lighigbm__new_model_mw_test_set_accuracy_all}, and still obtain powerful attacks.
 For 6,000 poisoned points, representing $1\%$ of the entire training set, the most effective strategy, \largeshapabs\ x \countshapabs, lowers \atksuccess\ on average to less than $3\%$.
Even at much lower poisoning rates ($0.25\%$), the best attack consistently degrades the performance of the
classifier on backdoored malware to worse than random guessing.
All the strategies induce small overall changes in the \bdrfp\, under $0.001$, with marginally larger increases correlated to larger poison sizes. 
We also observe minimal changes in \bdrgeneralize, on average below $0.1\%$.

Comparing the three attack strategies, we observe that the \indep\ attack composed by \largeshapabs\ and \countshapabs\ induces consistently high misclassification rates.  
It is also important to mention here that the \combined\ strategy is, as expected, remarkably stealthier.
We compared the accuracy of the clean model on the clean benign samples, against its accuracy of their respective backdoored counterparts, and observed very small differences across all attack runs.
In conclusion, we observe that the attack is extremely successful at inducing targeted mis-classification in the LightGBM
model, while maintaining good generalization on clean data, and low false positive rates.

\myparagraph{Targeting EmberNN.} Running the same series of attacks against EmberNN using the GradientSHAP explainer, we immediately notice that the Neural Network is generally more resilient to our attacks. 
Moreover, here the effect of trigger size is critical. 
\Cref{fig:embernn__new_model_mw_test_set_accuracy_all} shows the progression of accuracy loss over the watermarked malicious samples with the increase in trigger size, at a fixed $1\%$ poisoning rate. 
For example, under the most effective strategy, with a trigger size of 128 features, \atksuccess\ becomes
on average $0.75\%$, while \atksuccess\ averages $5.05\%$ at 32 features.
A critical element that distinguishes the three strategies on EmberNN, is the difference between the  accuracy of the clean model over the clean and backdoored benign samples.
While, the other tracked metrics show a behavior similar to the case of LightGBM, good generalization on clean data, with
\bdrgeneralize\ close to the original $99.11\%$ in most cases, and low false positives increase ($\approx 0.1-0.2\%$ average increase in \bdrfp), a clean EmberNN model often fails almost completely in recognizing backdoored benign points as goodware. 
Here, the \combined\ strategy emerges as a clear \say{winner,} being both very effective in inducing misclassification, and, simultaneously, minimizing the aforementioned difference, with an average absolute value of $\approx 0.3\%$.  
Interestingly, we also observed that the attack performance on the NN model is more strongly correlated with the size of the backdoor trigger than with the poison pool size, resulting in small (0.5\%) injection volumes inducing appreciable misclassification rates. 

\subsection{Limiting the Attacker}\label{eval_transfer_limited}

We consider here a \transferatk\ attacker without access to the model. 
This threat model prevents the attacker from being able to compute the SHAP values for the victim model, therefore, the backdoor has to be generated using a surrogate (or proxy) model sharing the same feature space. 
We simulated this scenario by attempting a backdoor transferability experiment between our
target models. 
Fixing the trigger size to 16 features we attacked LightGBM with a backdoor generated by the \combined\ strategy using the SHAP values extracted from an EmberNN surrogate model.
Then we repeated a similar procedure by creating a backdoor using the \indep\ strategy, with the combination of \largeshapabs\ and \countshapabs\ for feature and value selection respectively, computed on a LightGBM proxy, and used it to poison EmberNN's training set. 
The \atksuccess\ loss for both scenarios is shown in \Cref{fig:transfer_all_features}.  
The empirical evidence observed supports the conclusion that our attacks are transferable both ways.
In particular, we notice a very similar behavior in both models as we saw in the \baseatk\ scenario, with LightGBM being generally more susceptible to the induced misclassification. 
In that case, the trigger generated using the surrogate model produced a $\approx 82.3\%$ drop in accuracy on the backdoored malware set, for a poison size of $1\%$ of the training set.

Lastly, we evaluate the scenario in which the attacker has access to only a small subset of clean training data and uses the same model architecture as the victim (i.e., \limitedatk). 
We perform this experiment by training a LightGBM model with 20\% of the training data and using it to generate the trigger, which we then used to attack the LightGBM model trained over the entire dataset. 
Using the \indep\ strategy with \largeshapabs\ and \countshapabs\ over 16 features and a 1\% poison set size, we noticed
very little difference compared to the same attack where the SHAP values are computed over the entire training set ($\approx 4\%$ $\Delta$ \atksuccess).

\section{Problem-Space Considerations}\label{feasible}
\begin{figure*}[t!]
	\centering
	\begin{subfigure}[t]{0.5\textwidth}
		\centering
		
		\includegraphics[width=0.85\textwidth]{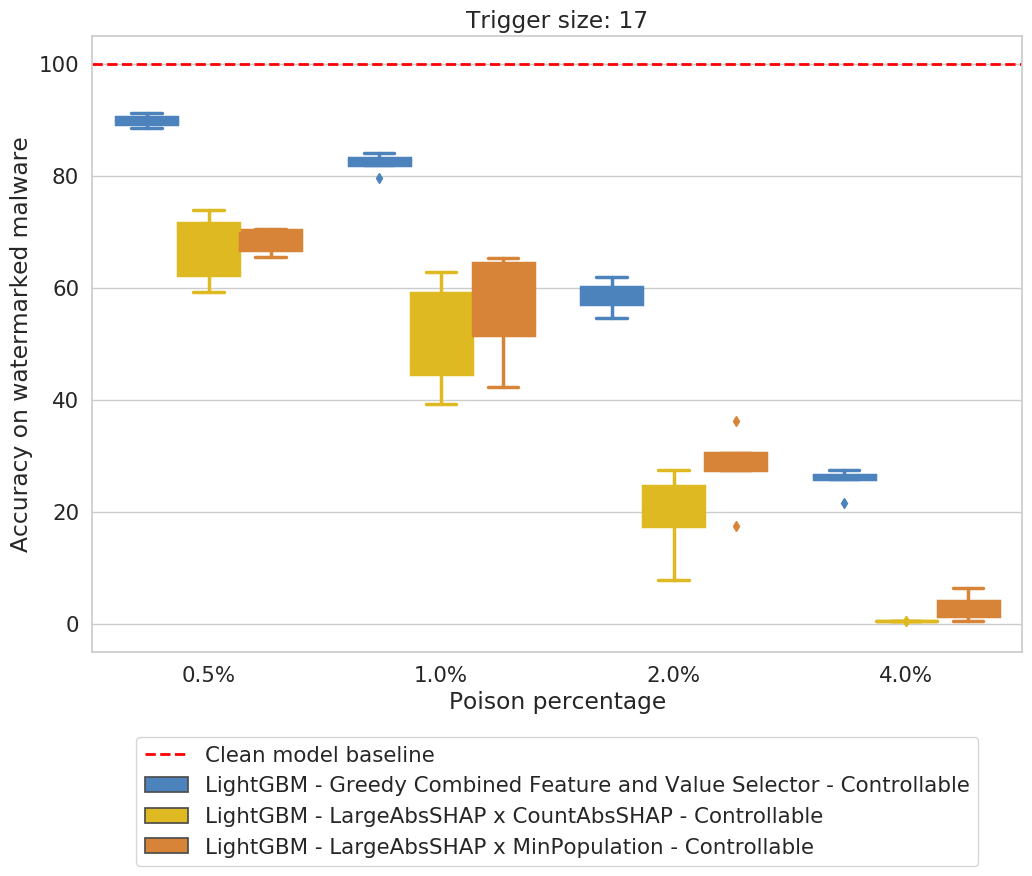}
		
		\caption{LightGBM target}
		
		\label{fig:lighigbm__new_model_mw_test_set_accuracy_feasible}
		
	\end{subfigure}%
	~~~ 
	\begin{subfigure}[t]{0.5\textwidth}
		\centering
		
		\includegraphics[width=0.85\textwidth]{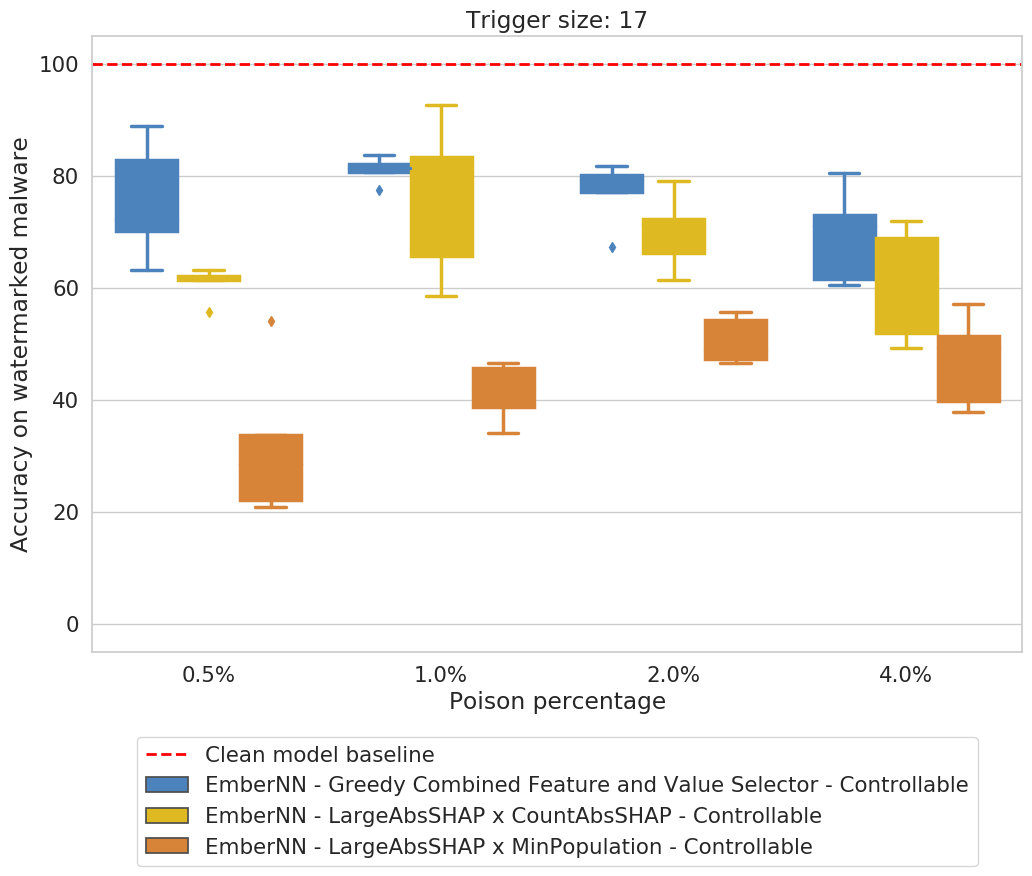}
		
		\caption{EmberNN target}
		
		\label{fig:embernn__new_model_mw_test_set_accuracy_feasible}
		
	\end{subfigure}
	
	\caption{Accuracy of the backdoor model over watermarked malicious samples. Lower \atksuccess\ is the
		result of stronger attacks. The watermark uses the subset of
		\numfeasible\ features of EMBER, modifiable by the \feasibleatk\ adversary.}
	
	\label{fig:new_model_test_set_accuracy_feasible}
	\vspace{0.15cm}
\end{figure*}

\begin{figure*}[t!]
	\centering
	\begin{subfigure}[t]{0.5\textwidth}
		\centering
		
		\includegraphics[width=0.85\textwidth]{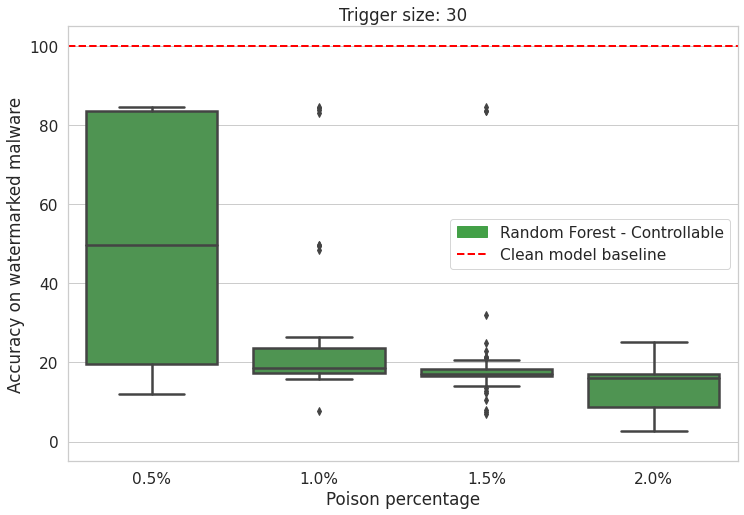}
		
		\caption{Random Forest classifier on Contagio data.}
		
		\label{fig:constrained_contagio}
		
	\end{subfigure}%
	~~~ 
	\begin{subfigure}[t]{0.5\textwidth}
		\centering
		
		\includegraphics[width=0.85\textwidth]{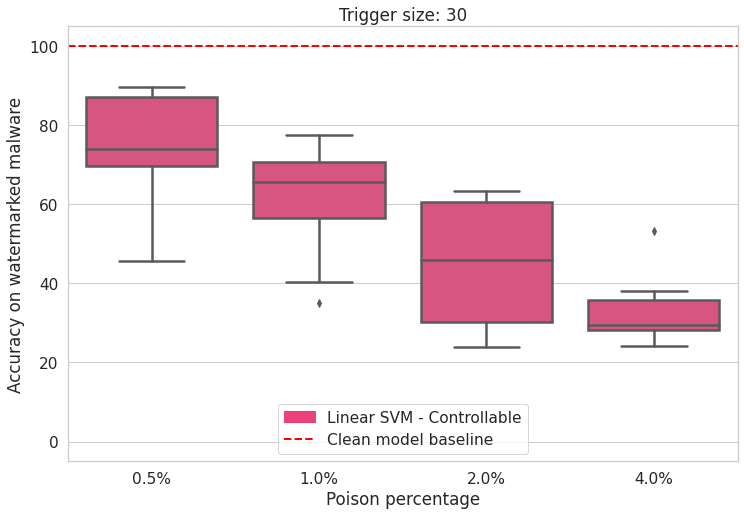}
		
		\caption{Linear SVM classifier on Drebin data.}
		
		\label{fig:constrained_transfer_drebin}
	\end{subfigure}
	
	\caption{50 attack runs for Contagio and 10 for Drebin, using the \combined\ strategy, with a 30-features trigger.}
	
	\label{fig:drebin_contagio}
\end{figure*}


In the previous section, we explored model-agnostic attack strategies when the attacker has full control of the features and can change their values at will.
A \feasibleatk\ attacker has to expend non-trivial effort to ensure that the backdoor generated in \emph{feature-space} does not break the semantics or otherwise compromise the functionality of binaries in the \emph{problem-space}~\cite{pierazzi_intriguing_2020}; that is backdoored goodware must maintain the original label and watermarked malware retain its malicious functionality.

\subsection{Windows PEs}

We implemented a backdooring utility using the \textit{pefile}~\cite{carrera_erocarrerapefile_nodate} library to create a generic tool that attempts to apply a given watermark to arbitrary Windows binaries.
Creating this utility in a sufficiently general way required specialized knowledge of the file structure for Windows Portable
Executable (PE) files, in particular when adding sections to the binaries.
Doing so required extending the section table with the appropriate sections, names, and characteristics, which in turn meant relocating structures that follow the section table, such as data directories and the sections themselves, to allow for arbitrary increases in the number of sections added.  

We also encountered several challenges that required us to drop certain features and consider dependencies among features that restrict the values they can take on.  
First, we realized that the vast majority of the features in EMBER \ignore{(2,316 of 2,351)} are
based on feature hashing, which is often used to vectorize arbitrarily large spaces into a fixed-length vector.  
For example, strings uncovered in the binary may be hashed into a small number of buckets to create a fixed-number of
counts.  
Given the preimage resistance of the hash function, directly manipulating these features by tampering with the binary would be extremely difficult, and consequently we discard all hash-based features, leaving us with just 35 directly-editable, non-hashed features. 
Next, we considered dependencies among the non-hashed features.  As it turns out,
many of the features are derived from the same underlying structures and
properties of the binary, and may result in conflicting watermarks that cannot
be simultaneously realized.  For example, the \textit{num\_sections} and
\textit{num\_write\_sections} features are related because each time we add a
writeable section, we necessarily increase the total number of sections.  To
handle these dependencies, we remove any features whose value is impacted by
more than one other feature (e.g., \textit{num\_sections}).  This allows us to
keep the maximal number of features without solving complex constraint
optimization problems.
The last challenge arose from the question of how to handle natural 
constraints of the problem space, such as cases where the watermark might require us to remove 
URLs or reduce the file size.  Here, the attacker has two choices: reduce the set of files that 
can be successfully watermarked or reduce the effectiveness of the watermark by adding constraints to the search algorithm that 
ensure maximal applicability, as shown in \Cref{method}.  Due to the large number of available Windows PE samples, we 
decided it was best for the attacker to sacrifice the samples, rather than lose attack effectiveness.  Later, we will show the 
opposite case for Android malware, where imposing constraints on the watermark was the preferable solution.

%


After reducing our set of features based on the above criteria, we are left
with 17 features that our generic watermarking utility can successfully
manipulate on arbitrary Windows binaries.  Examples of backdoor patterns can be found in \Cref{tab:feasibility_watermark},
\Cref{app:trigger}.  As we will see, despite the significant reduction in the
space of available features, our proposed attack strategies still show
significant effectiveness.
While developing the watermarking utility was
challenging, we believe it is well within the capabilities of a determined
attacker, and can subsequently be reused for a variety of attack campaigns.

\myparagraph{Attack Efficacy.} 
As shown in \Cref{fig:new_model_test_set_accuracy_feasible}, the effectiveness of the
attack is slightly decreased when the backdoor trigger is generated using
only the \numfeasible\ manipulable features supported by our watermarking
utility.  Such a \feasibleatk\ adversary, is, as expected, strictly less
powerful than the \baseatk\ attacker we explored in \Cref{eval}. On the other
hand, despite the strong limitations introduced to ease practical
implementation, we argue that the average accuracy loss is still extremely
relevant given the security critical application. Moreover, if we allow the
poison size to grow to $2\%$ of the overall training set, we obtain
\atksuccess\ levels comparable with the \baseatk\ at $1\%$ poison size on
LightGBM. 

To explore additional realistic scenarios, we combined the limitation over features control with lack of access to the original model, \feasibleatk\ - \transferatk.
As in \Cref{eval_transfer_limited}, we generated the watermark using a surrogate model, with the most effective \transferatk\ strategy we identified before, but this time restricted to the controllable features. 
We observed an average \atksuccess\ of 54.53\% and 56.76\% for LightGBM and EmberNN respectively. 
An even weaker and stealthier attacker could be obtained combining the characteristics of the previous adversary with a limited knowledge of the training data and the use of the \combined\ strategy. 
We evaluate the effect of this \feasibleatk\ - \transferatk - \limitedatk\ adversary, with a backdoor computed using an EmberNN surrogate, with access to only 20\% of the training set and applied to a LightGBM victim.  
Despite the extreme limitations imposed on the attacker, the effect on the model is still significant, with decreases in accuracy on points containing the trigger ranging from $\approx 10.8\%$ at 1\% poisoning, up to $\approx 40\%$ for a 4\% poisoning rate.

Lastly, we looked at the \feasibleatk\ - \kernelexpatk\ scenario, where we produced the SHAP values for only the manipulable features using the SHAP KernelExplainer, which operates purely by querying the model as a black-box. 
We target LightGBM, with the \largeshapabs\ x \countshapabs\ strategy, poisoning 1\% of the training set.
The resulting model exhibits an average \atksuccess\ of 44.62\%, which makes this attacker slightly weaker than one having access to model-specific SHAP explainers.
It is relevant to note here, that the adversary has to spend a significant amount of computation time to use the SHAP KernelExplainer.

\myparagraph{Behavior Preservation.}
We randomly selected the 100 goodware and 100 malware binaries from our dataset and poisoned each of them with the backdoor for the LightGBM and EmberNN models, resulting in a total of 200 watermarked binaries for each
model.
To determine the watermark effects on the binaries' functionality, we run each sample in a dynamic analysis sandbox, which uses a variety of static, dynamic, and behavioral analysis methods to determine whether a binary is malicious.  
This experiment helps evaluate three important aspects of our attack when
applied in the real world:  (i) the ability to keep the original labels on
watermarked goodware, (ii) the ability to maintain the original malicious
functionality of the watermarked malware, and (iii) the impact of semantic
restrictions on the features the adversary can use to carry out the poisoning.
The original and backdoored binaries were submitted to a dynamic analysis environment with an execution timeout of 120 seconds.
Table \ref{tab:feasibility_results}, in \Cref{app:trigger}, shows the results of our
experiments.  In the case of the LightGBM and EmberNN watermarks, both goodware
and malware have similar numbers of failed watermarking attempts due to the
physical constraints on the binaries, with the most prevalent reason (>90\%)
being binaries that were too large for the selected \textit{size} watermark.
For those files that were successfully watermarked, we observed that goodware
always maintained its original benign label, while malware retained its
malicious functionality in 61-66\% of the cases.  We also scanned our
watermarked binaries with ESET and Norton AntiVirus signature-based antivirus
engines, similar to those used by crowdsourced threat intelligence feeds, and
found that none of the goodware changed labels due to the presence of our
backdoor.  Overall, this indicates that an attacker could use up to 75\% of
the observed goodware and 47\% of the observed malware in these threat
intelligence feeds to launch their backdoor poisoning attack. This is
sufficient in real-world attacks as the adversary needs a small percentage of
poisoned binaries to execute the attack.
Finally, it is important to point out that our evaluation here focused on an adversary using commodity goodware and
malware. 
However, an advanced attacker may produce their own software to better align with the chosen watermark values and maximize the attack impact.

\subsection{Other Datasets}
\label{sec:otherdata}
PDF files and Android applications have been the object of a large body of research on malware classification and classifier evasion. Therefore, we focused on these two domains as examples for the adaptability of our explanation-based attack.

\myparagraph{PDF Files.} We worked with the Contagio\footnote{\url{http://contagiodump.blogspot.com/}} PDF data, consisting of 10,000 samples evenly distributed between benign and malicious, with 135-dimensional feature vectors extracted according to PDFRate~\cite{smutz_malicious_2012} specification.
To ensure our modifications were behavior-preserving, we developed a Python 3 port of the feature editor released\footnote{\url{https://github.com/srndic/mimicus}} with Mimicus~\cite{srndic_practical_2014}. This tool allowed us to parse the PDF files, apply the desired backdoor pattern, and read back a new feature vector after the poisoning to account for possible side effects, such as alterations in various size-based features.

Unfortunately, during our experimentation we ran into several bugs in the Mimicus feature editor that lead to inconsistent application of our otherwise valid watermark to the PDFs.  In particular, these issues forced us to reduce our trigger pattern to only 30 of the 35 features reported as 
modifiable in the paper, and to restrict our poisoning pool to only those files that were correctly backdoored.  Fixing these issues is beyond the scope of this work, but despite these limitations we were still able to poison enough samples to mount successful attacks.

\myparagraph{Android Applications.} In the Android domain, we used the well-studied Drebin~\cite{arp_drebin_2014} dataset containing 5,560 malicious and 123,453 benign apps, represented by Boolean vectors indicating which of the over 545,000 statically extracted features are present in the application. Such a large space of features is divided into 8 logical subsets, $S_1-S_4$ being characteristics of the Android \emph{manifest} file, and $S_5-S_8$ being extracted from the disassembled code.

To ensure no loss of functionality was inadvertently sustained as side effect of the trigger application, we borrowed the technique specified by Grosse et al.~\cite{grosse_adversarial_2016, grosse_adversarial_2017}. First, we restricted ourselves to only altering features belonging to subsets $S_1$ and $S_2$, representing the list of \emph{hardware components} and the list of \emph{permissions} requested by the application, respectively. Both these subsets belong to the manifest class of features and can be modified by changing a single line in the manifest file. Second, we forced our backdoor to be exclusively additive, meaning that no feature could be removed from an application as result of the poisoning. 

Other advanced (and computationally expensive) techniques may also be used to increase the number of manipulable features available to our attack strategy while still ensuring behavior preservation, such as \emph{organ harvesting}~\cite{pierazzi_intriguing_2020} for adversarial Android malware or behavioral \emph{oracles}~\cite{xu_automatically_2016} for PDF files. We believe that the improvement of feature-space to problem-space mapping methods, will greatly improve the effectiveness of explanation-guided poisoning attacks.

%
%
%
%
%

\myparagraph{Attack Efficacy.}
Having observed how our \combined\ strategy is both stealthy (more on this in \Cref{defense}), and especially adept at generating behavior preserving backdoors, we employed it for our experiments on the Contagio and Drebin datasets.
In both cases, we use the original model architecture proposed in the literature, therefore, we test our attack on a Random Forest classifier for the PDF files, and a Linear Support Vector Machine (SVM) classifier for the Android applications.

\Cref{fig:constrained_contagio} shows the reduction in accuracy of the poisoned Random Forest induced by our \feasibleatk\ adversary. It is interesting to observe that, probably due to the small size of the dataset combined with the necessity of limiting the poisoning pool to only the PDF files correctly modified by the editor utility, there appears to be a large amount of variance in the attack effectiveness at lower poison percentages. These effects fade away with larger poisoning pools. Overall, the attack is generally very successful, inducing, for instance, an average 21.09\% \atksuccess, at 1.5\% poisoning rate.

Applying the explanation attack to the Android data proved somewhat more challenging due to the sparsity of the feature space.
To handle the dimensionality issue, we first used L1 regularized logistic regression to select a subset of 991 features, then we trained a surrogate LightGBM mode and used the surrogate to compute the SHAP values. 
This corresponds to a \transferatk-\feasibleatk\ adversary. 
A 30-feature backdoor thus computed was then applied to the original 545K-dimensional vectors used to train the Linear SVM.
\Cref{fig:constrained_transfer_drebin} shows the effect of the poisoning on the accuracy of the model on backdoored malware.
For instance, at 2\% poisoning rate, the attack lowers the model accuracy on backdoored samples to 42.9\% on average
We also observed minimal loss of \bdrgeneralize\, within 0.03\%, and change in \bdrfp, less than 0.08\%, on average.

\section{Mitigation}\label{defense}
\begin{table*}[t!]
\def\arraystretch{1.05}
\scriptsize
\centering
\begin{tabular}{c c c c c c c c}
\toprule
\textbf{Target} & \textbf{Strategy} &  \textbf{\atksuccess} & \textbf{Mitigation} & \textbf{New \atksuccess} & \textbf{Poisons} & \textbf{Goodware} \\
& & \textbf{(after attack)} & & \textbf{(after defense)} & \textbf{Removed} & \textbf{Removed}\\
\midrule
\multirow{9}{*}{LightGBM} & \multirow{3}{*}{\makecell{LargeAbsSHAP x \\ MinPopulation}} & \multirow{3}{*}{0.5935} &             HDBSCAN & 0.7422 &  3825 & 102251 \\
 & & &  Spectral Signature & 0.7119 &   962 & 45000\\
 & & &  Isolation Forest & 0.9917 & 6000 & 11184\\
\cline{4-7}
 & \multirow{3}{*}{\makecell{LargeAbsSHAP x \\ CountAbsSHAP}} & \multirow{3}{*}{0.5580} &             HDBSCAN & 0.7055 &  3372 & 93430 \\
 & & &  Spectral Signature & 0.6677 &   961 & 44999\\
 & & &  Isolation Forest & 0.9921 & 6000 & 11480\\
\cline{4-7}
 & \multirow{3}{*}{\makecell{Combined Feature \\ Value Selector}} & \multirow{3}{*}{0.8320} &             HDBSCAN & 0.8427 &  1607 & 115282 \\
 & & &  Spectral Signature & 0.7931 &   328 & 45000\\
 & & &  Isolation Forest & 0.8368 & 204 & 8927\\
\midrule
\multirow{9}{*}{EmberNN} & \multirow{3}{*}{\makecell{LargeAbsSHAP x \\ MinPopulation}}  & \multirow{3}{*}{0.4099} &             HDBSCAN & 0.3508 &  3075 & 137597 \\
 & & &  Spectral Signature & 0.6408 &   906 & 45000\\
 & & &  Isolation Forest & 0.9999 & 6000 & 14512 \\
\cline{4-7}
 & \multirow{3}{*}{\makecell{LargeAbsSHAP x \\ CountAbsSHAP}}  & \multirow{3}{*}{0.8340} &             HDBSCAN & 0.5854 &  2499 & 125460 \\
 & & &  Spectral Signature & 0.8631 &   906 & 45000\\
 & & &  Isolation Forest & 0.9999 & 6000 & 15362 \\
\cline{4-7}
 & \multirow{3}{*}{\makecell{Combined Feature \\ Value Selector}}  & \multirow{3}{*}{0.8457} &             HDBSCAN & 0.8950 &  1610 & 120401 \\
 & & &  Spectral Signature & 0.9689 &   904 & 45000\\
 & & &  Isolation Forest & 0.8030 & 175 & 13289 \\
\bottomrule
\end{tabular}

\caption{Mitigation results for both LightGBM and EmberNN. All attacks were
    targeted towards the \numfeasible\ controllable features (see
    \Cref{feasible}), with a $1\%$ poison set size, 6000 backdoored benign
    samples. We show \atksuccess\ for the backdoored model, and after the
defense is applied. We also include number of poisoned and goodware  points
filtered out by the defensive approaches.}

\label{tab:mitigations}
\end{table*}

\ignore{Recently, researchers started tackling the problem of defending against
backdoor
attacks~\cite{chen_detecting_2018,tran_spectral_2018,liu_fine-pruning_2018,wang_neural_2019}.
Nearly all existing defensive approaches, however, are specifically targeted at
computer vision deep neural network models, and assume threat models in which
the adversary has actively tampered with the training labels.  These
limitations make them hard to adapt to the class of model-agnostic, clean-label
attacks we are interested in. We discuss here a few representative examples of
the currently published works on the subject.  }

Recently, researchers started tackling the problem of defending against
backdoor
attacks~\cite{chen_detecting_2018,tran_spectral_2018,liu_fine-pruning_2018,wang_neural_2019}.
Nearly all existing defensive approaches, however, are specifically targeted at
computer vision Deep Neural Networks, and assume 
 adversaries that actively tamper with the training labels.  These
limitations make them hard to adapt to the class of model-agnostic, clean-label
attacks we are interested in. We discuss here representative related work.


Tran et al.~\cite{tran_spectral_2018} propose a defensive method based on
\emph{spectral signatures}, which relies on detecting two
$\epsilon$-spectrally separable subpopulations based on SVD decomposition.
 Chen et al.~\cite{chen_detecting_2018}  rely on the representation learned by the CNN and perform k-means clustering on the activations of the last convolutional layer.  The defense of Liu et al.~\cite{liu_fine-pruning_2018} is based on combining network fine tuning and neuron pruning, making it specific to neural networks.
 Finally, NeuralCleanse~\cite{wang_neural_2019} is based on the intuition that in a backdoored model, the perturbation necessary to induce a misclassification towards the targeted class should be smaller than that required to obtain different labels. This approach was designed considering multi-class classification problem, as encountered in image recognition, and the suggested filtering and pruning mitigation are neural-network specific.

\ignore{Wang et al.~\cite{wang_neural_2019} introduce a
defensive procedure called NeuralCleanse based on the intuition that in a
backdoored model, the perturbation necessary to induce a misclassification
towards the targeted class should be smaller than that required to obtain
different labels. This approach was designed considering multi-class
classification problem, as encountered in image recognition, moreover the
filtering and pruning mitigation suggested directly operate on the second to
last convolutional layer neuron representation. }



\myparagraph{Considered Defensive Approaches.} 
According to our threat model, the defender is assumed to:
\begin{enumerate*}[label=(\roman*)]

    \item have access to the (poisoned) training data;
        
    \item have access to a small set of clean labeled data. This common
        assumption in adversarial ML fits nicely with the context since
        security companies often have access to internal, trusted, data
        sources; and
        
    \item know that the adversary will target the most relevant features. 

\end{enumerate*}

\ignore{
Our first intuition in designing a mitigation was to use the \clnfp\ drop as an
alarm signaling that a dissemination campaign was poisoning the training data.
This basic approach, however, presents multiple shortcomings.
\begin{enumerate*}[label=(\roman*)]

    \item It fails when the attacker employs stealthier strategies, such as
        \combined.

    \item It cannot distinguish between adversarially crafted poison samples
        and unseen-before samples. Therefore, the accuracy drop induced by a
        small poison size, such as $1-2\%$, can be easily confused with a
        natural drop in accuracy due to previously unseen data subpopulations.

    \item Even in the more optimistic scenarios, it can provide only an alert,
        being unable to identify the actual backdoored points.
\end{enumerate*}
}


We evaluate three mitigation strategies over a reduced feature space obtained by selecting a fixed number (32) of the most important features.
First, a state-of-the-art defensive strategy, spectral
    signatures~\cite{tran_spectral_2018}, which we adapt by computing the
    singular value decomposition of the benign samples over the new feature
    space. Then, as in the original paper, we compute the \emph{outlier score}
    by multiplying the top right singular vector and we filter out the samples
    with the highest 15\% scores.
Second, hierarchical density-based clustering,
    (HDBSCAN)~\cite{campello_density-based_2013}, inspired by Chen et
    al's~\cite{chen_detecting_2018} use of k-means for defensive clustering over
    neuron activations.  We borrow the idea, using HDBSCAN instead, with the
    intuition that watermarked samples form a subspace of high density in
    the reduced feature space, and generate a tight cluster.  Additionally,
    HDBSCAN does not require a fixed number of clusters, but has two other
    parameters that control the cluster density  (minimum size
    of a cluster, set at 1\% of the training benign data, 3000 points, and
    minimum number of samples to form a dense region, set at 0.5\%, 600
    points).  As in~\cite{chen_detecting_2018}, we compute Silhouette scores on
    the resulting clusters, to obtain an estimate of the intra-cluster
    similarity of a sample compared to points from its nearest neighboring
    cluster, and filter out samples from each cluster with a probability
    related to the cluster silhouette score.
Third, isolation forest~\cite{liu_isolation_2008}, an algorithm for unsupervised
    anomaly detection based on identifying rare and different points instead
    of building a model of a normal sample. The intuition here is that such an
    anomaly detection approach might identify the watermarked samples as
    outliers due to their similarity compared to the very diverse background
    points. We experiment with default parameters of Isolation Forest.

\myparagraph{Results of Mitigation Strategies.} \Cref{tab:mitigations} shows
the effect of these three mitigation strategies over the different models and
attack strategies. Two main takeaways emerge from these empirical results.
First, the Isolation Forest, trained on the reduced feature space, is often
capable of correctly isolating all the backdoored points with relatively low
false positives. Note that this happens exclusively when an Isolation Forest is
trained on the transformed dataset (reduced to most important features). The
same algorithm applied in the original feature space detects only a tiny
fraction of the backdoored points ($\approx 1\%$), with similar results obtained also on Drebin (0\%) and Contagio (12.5\%), thus reinforcing the observation in~\cite{tran_spectral_2018} that the subpopulations are not
sufficiently separable in the original feature space. Second, none of the
mitigation approaches was able to isolate the points attacked with watermarks
produced with the \combined\ strategy on PE files. This confirms that the \combined\ attack
strategy is much more stealthy compared to both \indep\ strategies.

We note that the proposed mitigations are only a first practical step in
defending against clean-label backdoor attacks in a model-agnostic setting. We leave a deeper investigations of more general defensive methods, as a topic of future work. Protecting ML systems from adversarial attacks is an intrinsically hard problem~\cite{carlini_evaluating_2019}. We argue that defending against our backdoor attacks is extremely challenging due to the
combined effect of the small subpopulation separability induced by clean-label attacks, and the difficulty of distinguishing dense  regions generated by the attack from other dense regions naturally occurring in diverse sets of benign binaries.



\section{Related Work}\label{relwork}

An early line of research introduced by Perdisci et al.~\cite{perdisci_misleading_2006} and Newsome et al.~\cite{newsome_paragraph_2006} demonstrated methods for 
polluting automated polymorphic worm detectors such as Polygraph~\cite{newsome_polygraph_2005}. The first~\cite{perdisci_misleading_2006} introduced purposely crafted noise in the traces used for signature generation  to prevent the generation of useful signatures; the second~\cite{newsome_paragraph_2006} proposed \emph{red herring} attacks, where the goal of the adversary is to force the generated system to rely on spurious features for classification, which will then be excluded from the evading sample. Red herring attacks are particularly interesting for us, being the first to suggest that an adversary does not necessarily need control over data labels in order to cause failures in the downstream classifier, thus foreshadowing \emph{clean-label} poisoning. Successive work by Venkataraman et al.~\cite{venkataraman_limits_2008} generalizes these results by providing lower bounds on the number of mistakes made by a signature generation algorithm based on conjunctions of boolean features. Theoretical bounds on poisoning attacks against an online centroid anomaly detection method have subsequently been analyzed by Kloft and Laskov~\cite{pmlr-v9-kloft10a} in the context of network intrusion detection.
Concurrently, researchers started to analyze possible countermeasures to poisoning attempts against anomaly detection systems deployed to discover abnormal patterns in network traces. Cretu et al.~\cite{cretu_casting_2008} developed a methodology to sanitize training data based on the output of an ensemble of micro models, trained on small portions of the data, combined through simple voting schemes. Rubinstein et al.~\cite{rubinstein_antidote_2009} later proposed to leverage methods from robust statistics to minimize the effect of small poison quantities on network traffic anomaly detectors based on Principal Component Analysis.

More recent research by Biggio et al.~\cite{biggio_poisoning_2012} brought to light the problem of poisoning attacks against modern machine learning models by proposing an availability attack based on gradient ascent against support vector machines.
Successive work~\cite{biggio_poisoning_2014}, demonstrated the relevance of ML poisoning in the domain of malware classification by targeting Malheur~\cite{rieck_automatic_2011}, a malware behavioral clustering tool.
Later research by Xiao et al.~\cite{xiao_is_2015} showed that feature selection methods, like LASSO,
ridge regression, and elastic net, were susceptible to small poison sizes.
Gradient-based poisoning availability attacks have been shown against regression~\cite{Jagielski18} and neural
networks~\cite{munoz-gonzalez_towards_2017}, and the transferability of these attacks has been demonstrated~\cite{Demontis19}. Recently, Suciu et al.~\cite{suciu_when_2018} proposed a framework for defining attacker models in the poisoning space, and developed StingRay, a multi-model target poisoning attack methodology.

\ignore{Biggio et al.~\cite{biggio_poisoning_2012} was one of the first to
    bring the problem of poisoning attacks against ML models to light.  In this
    paper the authors proposed a gradient ascent based modification of the
    input data aimed at increasing the testing error of a Support Vector
    Machine classifier.  Successive work~\cite{biggio_poisoning_2014},
    demonstrated the relevance of poisoning attacks in the computer security
    domain by attacking a renown malware behavioral clustering tool,
    Malheur\cite{rieck_automatic_2011}.  Later research by Xiao et
    al.~\cite{xiao_is_2015} showed that feature selection methods, like LASSO,
    ridge regression, and elastic net, were susceptible even to relatively
    small poison sizes, $~5\%$.  Luis Muñoz-González et
    al.~\cite{munoz-gonzalez_towards_2017} experimented with
    \emph{availability} attacks, which maximize the classification error of the
    victim model, on a set of ransomware samples, and first explored the
    transferability property of poisoning attacks.  Recently, Suciu et
al.~\cite{suciu_when_2018} proposed a framework for defining attacker models in
the poisoning space, and developed StingRay, a multi-model target poisoning
attack methodology.}

Backdoor attacks were introduced by Gu et al. in BadNets~\cite{gu_badnets:_2017}, identifying a supply chain vulnerability in modern machine learning as-a-service pipelines.  
Liu et al.~\cite{liu_trojaning_2018} explored introducing trojan triggers in image recognition Neural Networks, without requiring access to the original training data, by partially re-training the models.  
Later works by Turner et al.~\cite{turner_clean-label_2019} and Shafahi et al.~\cite{shafahi_poison_2018} further improved over the existing attacks by devising clean-label strategies.

\ignore{Backdoor attacks were first introduced by Gu et al. in
    BadNets~\cite{gu_badnets:_2017}, in which the authors identified a supply
    chain vulnerability in modern machine learning as-a-service pipelines.  An
    attacker with complete access to the training procedure could induce the
    model to memorize a specific pixel pattern injected in the training images,
    and associate it to a class of the attacker's choice. During inference, the
    same pattern could be injected in a test image in order to trigger the
    desired classification.  \cite{liu_trojaning_2018} by Liu et al., explored
    introducing trojan triggers in image recognition Neural Networks, without
    requiring access to the original training data, by partially re-training
    the models.  Later works by Turner et al.~\cite{turner_clean-label_2019}
    and Shafahi et al.~\cite{shafahi_poison_2018} further improved over the
    existing attacks by devising strategies to allow the attacker to inject a
    backdoor without actively controlling the data labeling process.  To the
    best of our knowledge, no previous work has focused on backdoor attacks
    aimed specifically at malicious software classifiers. Moreover, we are not
    aware of any attempt at injecting backdoors in Gradient Boosting models.}

\section{Discussion and Conclusion}\label{discuss}
With this work we begin shedding light on new ways of implementing clean-label backdoor attacks, a
threat vector that we believe will only grow in relevance in the coming
years. We showed how to conduct backdoor poisoning attacks that are
model-agnostic, do not assume control over the labeling process, and can be
adapted to very restrictive adversarial models. For instance, an attacker with
the sole knowledge of the feature space can mount a realistic attack by
injecting a relatively small pool of poisoned samples (1\% of training set) and
induce high misclassification rates in backdoored malware samples.
Additionally, we designed the \combined\ strategy that creates backdoored points in
high-density regions of the legitimate samples, making it very difficult to
detect with common defenses.
Based on our exploration of these attacks, we believe explanation-guided attack strategies could also be applicable to other feature-based models, outside of the security domain.

Finally, there are some limitations of this work that we would like to expose.
First, the attacks we explored rely on the
attacker knowing the feature space used by the victim model. While this
assumption is partially justified by the presence of natural features in the
structure of executable files, we consider the development of more generic
attack methodologies, which do not rely on any knowledge from the adversary's
side, as an interesting future research direction. Second,
designing a general mitigation method, particularly against our stealthy
\combined\ attack strategy, remains a challenging problem for future
work. Lastly, adaptation of these attacks to other malware classification
problems that might rely on combining static and dynamic analysis is also a
topic of future investigation. 



\section*{Acknowledgments}
We would like to thank Jeff Johns for his detailed feedback on a draft of this paper and many discussions on backdoor poisoning attacks, and the anonymous reviewers for their insightful comments and valuable suggestions. 
We thank FireEye for sponsoring research done at Northeastern University for this project. 
The work done at Northeastern University was also partially sponsored by the U.S. Army Combat Capabilities Development Command Army Research Laboratory and was accomplished under Cooperative Agreement Number W911NF-13-2-0045 (ARL Cyber Security CRA).
The views and conclusions contained in this document are those of the authors and should not be interpreted as representing the official policies, either expressed  or implied, of the Combat Capabilities Development Command Army Research Laboratory or the U.S. Government.  
The U.S. Government is authorized to reproduce and distribute reprints for Government purposes notwithstanding any copyright notation


\Urlmuskip=0mu plus 1mu
\bibliographystyle{plain}
\bibliography{prj_malw_bdr}

\FloatBarrier
\appendix
\section{Additional Results}\label{appendix_results}
\begin{table}[h]
	\footnotesize
	\centering
	\begin{tabular}{ l | c | c  } 
		\toprule
		\textbf{Feature} & \textbf{LightGBM} & \textbf{EmberNN} \\
		\midrule
		major\_image\_version & 1704 & 14 \\
		major\_linker\_version & 15 & 13 \\
		major\_operating\_system\_version & 38078 & 8 \\
		minor\_image\_version & 1506 & 12 \\
		minor\_linker\_version & 15 & 6 \\
		minor\_operating\_system\_version & 5 & 4 \\
		minor\_subsystem\_version & 5 & 20 \\
		MZ\_count & 626 & 384 \\
		num\_read\_and\_execute\_sections & 20 & 66 \\
		num\_unnamed\_sections & 11 & 6 \\
		num\_write\_sections & 41 & 66 \\
		num\_zero\_size\_sections & 17 & 17 \\
		paths\_count & 229 & 18 \\
		registry\_count & 0 & 33 \\
		size & 1202385 & 817664 \\
		timestamp & 1315281300 & 1479206400 \\
		urls\_count & 279 & 141 \\
		\bottomrule
	\end{tabular}
	\caption{Watermarks for LightGBM and EmberNN used during feasibility testing.\\}
	\label{tab:feasibility_watermark}
\end{table}

\begin{table}[]
	\footnotesize
	\centering
	\begin{tabular}{ c | c | c | c } 
		\toprule
		\textbf{Dataset} & \textbf{Label} & \textbf{Result} & \textbf{Count}\\
		\midrule
		\multirow{4}{*}{Original} & \multirow{2}{*}{Goodware} & Dynamic Benign & 100 \\
		&  & Dynamic Malicious & 0 \\
		\cline{2-4}
		&  \multirow{2}{*}{Malware} & Dynamic Benign & 7 \\
		&  & Dynamic Malicious & 93 \\
		\hline
		\multirow{6}{*}{LightGBM} & \multirow{3}{*}{Goodware} & Failed & 25 \\
		&  & \textbf{Dynamic Benign} & \textbf{75} \\
		&  & Dynamic Malicious & 0 \\
		\cline{2-4}
		& \multirow{3}{*}{Malware} & Failed & 23 \\
		&  & Dynamic Benign & 30 \\
		&  & \textbf{Dynamic Malicious} & \textbf{47} \\
		\hline
		\multirow{6}{*}{EmberNN} & \multirow{3}{*}{Goodware} & Failed & 33 \\
		&  & \textbf{Dynamic Benign} & \textbf{67} \\
		&  & Dynamic Malicious & 0 \\
		\cline{2-4}
		& \multirow{3}{*}{Malware} & Failed & 33 \\
		&  & Dynamic Benign & 23 \\
		&  & \textbf{Dynamic Malicious} & \textbf{44} \\
		\bottomrule
	\end{tabular}
	\caption{Summary of results analyzing a random sample of 100 watermarked
		goodware and malware samples in the dynamic analysis environment.}
	\label{tab:feasibility_results}
\end{table}

Here the reader will find additional details on the experimental results and
feature analysis that help providing a general idea on the effectiveness and
feasibility of the studied attacks.

\subsection{Attack Results}\label{app:attacks}

\Cref{tab:largecount}, \Cref{tab:largemin}, and \Cref{tab:greedy} report
additional experimental results for the multiple runs of the attack with
different strategies.  All the attacks were repeated for 5 times and the tables
report average results.

\subsection{Feasible Backdoor Trigger}\label{app:trigger}

With our watermarking utility we were able to control \numfeasible\ features
with relative ease.  \Cref{tab:feasibility_watermark} shows the feature-value
mappings for two example backdoor triggers computed on the LightGBM and EmberNN
models, which we fed to the static and dynamic analyzers to gauge the level of
label retention after the adversarial modification.
\Cref{tab:feasibility_results} summarizes the results of the dynamic analyzer
over 100 randomly sampled benign and malicious executables from the EMBER
dataset.

\begin{table*}
	\centering
	\footnotesize
	\caption{LargeAbsSHAP x CountAbsSHAP - All features. Average percentage over 5 runs.}
	\begin{tabularx}{0.5\textwidth}{XXXXX}
		\toprule
		Trigger Size &  Poisoned Points & \atksuccess & \bdrgeneralize & \bdrfp \\
		\midrule
		4 &        1500 & 65.8713 & 98.6069 & 0.0114  \\
		4 &        3000 & 55.8789 & 98.5995 & 0.0116  \\
		4 &        6000 & 40.3358 & 98.6081 & 0.0116  \\
		4 &       12000 & 20.1088 & 98.6060 & 0.0118  \\
		\hline
		8 &        1500 & 30.8596 & 98.6335 & 0.0114  \\
		8 &        3000 & 10.1038 & 98.6212 & 0.0115  \\
		8 &        6000 &  2.8231 & 98.6185 & 0.0116  \\
		8 &       12000 &  0.0439 & 98.5975 & 0.0121  \\
		\hline
		16 &        1500 &  2.4942 & 98.6379 & 0.0114  \\
		16 &        3000 &  0.9899 & 98.6185 & 0.0114  \\
		16 &        6000 &  0.0205 & 98.5948 & 0.0116  \\
		16 &       12000 &  0.0138 & 98.6323 & 0.0117  \\
		\bottomrule
		\multicolumn{5}{c}{}\\
		\multicolumn{5}{c}{LightGBM}\\
	\end{tabularx}%
	~~~~~~
	\begin{tabularx}{0.5\textwidth}{XXXXX}
		\toprule
		Trigger Size &  Poisoned Points & \atksuccess & \bdrgeneralize & \bdrfp \\
		\midrule
		16 &        3000 & 21.0122 & 99.0832 & 0.0073 \\
		16 &        6000 & 36.7591 & 99.0499 & 0.0082  \\
		16 &       12000 & 53.8470 & 99.0729 & 0.0079  \\
		\hline
		32 &        3000 & 13.2336 & 99.0608 & 0.0078  \\
		32 &        6000 & 20.3952 & 99.1152 & 0.0070  \\
		32 &       12000 & 28.3413 & 99.0856 & 0.0074  \\
		\hline
		64 &        3000 &  5.8046 & 99.0723 & 0.0084  \\
		64 &        6000 & 11.1986 & 99.0959 & 0.0078  \\
		64 &       12000 & 11.5547 & 99.0998 & 0.0070  \\
		\hline
		128 &        3000 &  2.4067 & 99.0810 & 0.0075  \\
		128 &        6000 &  1.6841 & 99.0688 & 0.0075  \\
		128 &       12000 &  2.8298 & 99.1088 & 0.0074  \\
		\bottomrule
		\multicolumn{5}{c}{}\\
		\multicolumn{5}{c}{EmberNN}\\
	\end{tabularx}
	\label{tab:largecount}
\end{table*}

\begin{table*}
	\centering
	\footnotesize
	\caption{LargeAbsSHAP x MinPopulation - All features. Average percentage over 5 runs.}
	\begin{tabularx}{0.5\textwidth}{XXXXX}
		\toprule
		Trigger Size &  Poisoned Points & \atksuccess & \bdrgeneralize & \bdrfp \\
		\midrule
		4 &        1500 & 62.3211 & 98.5985 & 0.0115  \\
		4 &        3000 & 52.5933 & 98.6144 & 0.0114 \\
		4 &        6000 & 30.8696 & 98.6044 & 0.0116  \\
		4 &       12000 & 20.3445 & 98.5836 & 0.0118  \\
		\hline
		8 &        1500 & 32.0446 & 98.6128 & 0.0114  \\
		8 &        3000 & 20.5850 & 98.6159 & 0.0115  \\
		8 &        6000 & 14.9360 & 98.6087 & 0.0115 \\
		8 &       12000 &  1.9214 & 98.6037 & 0.0117  \\
		\hline
		16 &        1500 &  4.3328 & 98.6347 & 0.0114  \\
		16 &        3000 &  1.4490 & 98.6073 & 0.0115  \\
		16 &        6000 &  0.1670 & 98.6301 & 0.0115  \\
		16 &       12000 &  0.0026 & 98.6169 & 0.0118 \\
		\bottomrule
		\multicolumn{5}{c}{}\\
		\multicolumn{5}{c}{LightGBM}\\
	\end{tabularx}%
	~~~~~
	\begin{tabularx}{0.5\textwidth}{XXXXX}
		\toprule
		Trigger Size &  Poisoned Points & \atksuccess & \bdrgeneralize & \bdrfp \\
		\midrule
		16 &        3000 & 18.8691 & 99.1219 & 0.0074  \\
		16 &        6000 & 33.5211 & 99.0958 & 0.0079  \\
		16 &       12000 & 50.6499 & 99.0942 & 0.0080  \\
		\hline
		32 &        3000 &  9.1183 & 99.1189 & 0.0075  \\
		32 &        6000 & 12.1103 & 99.0827 & 0.0078  \\
		32 &       12000 & 14.6766 & 99.1127 & 0.0071 \\
		\hline
		64 &        3000 &  3.4980 & 99.1170 & 0.0075  \\
		64 &        6000 &  6.2418 & 99.1234 & 0.0072  \\
		64 &       12000 &  6.8627 & 99.0941 & 0.0075  \\
		\hline
		128 &        3000 &  0.9514 & 99.0675 & 0.0082  \\
		128 &        6000 &  1.6012 & 99.0824 & 0.0082  \\
		128 &       12000 &  1.6200 & 99.0816 & 0.0074  \\
		\bottomrule
		\multicolumn{5}{c}{}\\
		\multicolumn{5}{c}{EmberNN}\\
	\end{tabularx}
	\label{tab:largemin}
\end{table*}

\begin{table*}
	\centering
	\footnotesize
	\caption{Greedy Combined Feature and Value Selector - All features. Average percentage over 5 runs.}
	\begin{tabularx}{0.5\textwidth}{XXXXX}
		\toprule
		Trigger Size &  Poisoned Points & \atksuccess & \bdrgeneralize & \bdrfp \\
		\midrule
		4 &        1500 & 63.3370 & 98.5976 & 0.0113 \\
		4 &        3000 & 60.6706 & 98.6320 & 0.0114  \\
		4 &        6000 & 54.3283 & 98.6211 & 0.0114 \\
		4 &       12000 & 40.2437 & 98.6099 & 0.0118  \\
		\hline
		8 &        1500 & 49.5246 & 98.6290 & 0.0113  \\
		8 &        3000 & 37.3295 & 98.6153 & 0.0113  \\
		8 &        6000 & 23.6785 & 98.6147 & 0.0117  \\
		8 &       12000 & 17.7914 & 98.6282 & 0.0117  \\
		\hline
		16 &        1500 &  0.8105 & 98.6195 & 0.0113  \\
		16 &        3000 &  0.6968 & 98.6170 & 0.0115  \\
		16 &        6000 &  0.0565 & 98.6241 & 0.0116 \\
		16 &       12000 &  0.0329 & 98.6173 & 0.0118  \\
		\bottomrule
		\multicolumn{5}{c}{}\\
		\multicolumn{5}{c}{LightGBM}\\
	\end{tabularx}%
	~~~~~
	\begin{tabularx}{0.5\textwidth}{XXXXX}
		\toprule
		Trigger Size &  Poisoned Points & \atksuccess & \bdrgeneralize & \bdrfp \\
		\midrule
		16 &        3000 & 11.6613 & 99.1014 & 0.0082  \\
		16 &        6000 & 11.0876 & 99.1105 & 0.0078 \\
		16 &       12000 & 10.5981 & 99.0958 & 0.0079 \\
		\hline
		32 &        3000 &  4.8025 & 99.0747 & 0.0087\\
		32 &        6000 &  5.0524 & 99.1167 & 0.0082  \\
		32 &       12000 &  4.4665 & 99.1335 & 0.0072  \\
		\hline
		64 &        3000 &  1.9074 & 99.1012 & 0.0076  \\
		64 &        6000 &  1.8246 & 99.0989 & 0.0077  \\
		64 &       12000 &  1.8364 & 99.1117 & 0.0071  \\
		\hline
		128 &        3000 &  0.7356 & 99.0926 & 0.0082  \\
		128 &        6000 &  0.7596 & 99.1219 & 0.0080\\
		128 &       12000 &  0.7586 & 99.1014 & 0.0072\\
		\bottomrule
		\multicolumn{5}{c}{}\\
		\multicolumn{5}{c}{EmberNN}\\
	\end{tabularx}
	\label{tab:greedy}
\end{table*}

\end{document}